\documentclass{article}

\usepackage{microtype}
\usepackage{graphicx}
\usepackage{subfigure}
\usepackage{booktabs} 
\usepackage{multirow}
\usepackage{amssymb, amsmath}
\usepackage{array}
\usepackage{amsthm}
\newtheorem{proposition}{Proposition}
\usepackage{smile}
\usepackage{siunitx}
\usepackage{xspace}
\usepackage{float}
\usepackage{soul}
\usepackage{colortbl}
\definecolor{LightCyan}{rgb}{0.8, 0.9, 1}
\definecolor{my_gray}{rgb}{0.86, 0.86, 0.86}
\definecolor{Highlight}{rgb}{0.89,0.89,0.94}

\usepackage{hyperref}
\usepackage{enumitem}

\usepackage[textsize=tiny]{todonotes}
\newcommand{\yiqun}[1]{\textcolor{purple}{#1}}

\newcommand{\model}{\textsc{ConfDiff}\xspace}

\usepackage[accepted]{icml2024}

\usepackage[compact]{titlesec}
\usepackage{sidecap}
\titlespacing*{\section}{0pt}{*0.3}{*0.3}
\titlespacing*{\subsection}{0pt}{*0.3}{*0.3}
\titlespacing*{\subsubsection}{0pt}{*0.3}{*0.3}

\icmltitlerunning{Protein Conformation Generation via Force-Guided SE(3) Diffusion Models}

\begin{document}

\twocolumn[
\icmltitle{Protein Conformation Generation via Force-Guided $\mathrm{SE}(3)$ Diffusion Models}



\icmlsetsymbol{equal}{*}

\begin{icmlauthorlist}
\icmlauthor{Yan Wang}{equal,goo,to}
\icmlauthor{Lihao Wang}{equal,goo}
\icmlauthor{Yuning Shen}{goo}
\icmlauthor{Yiqun Wang}{goo}
\icmlauthor{Huizhuo Yuan}{ed}
\icmlauthor{Yue Wu}{ed}
\icmlauthor{Quanquan Gu}{goo}
\end{icmlauthorlist}

\icmlaffiliation{to}{School of Mathematical Sciences, Tongji University, Shanghai (this work was done during Yan’s internship at
ByteDance Research)}
\icmlaffiliation{goo}{ByteDance Research}
\icmlaffiliation{ed}{Department of Computer Science, University of California, Los Angeles (this work was done during Huizhuo and Yue’s internship at
ByteDance Research)}

\icmlcorrespondingauthor{Quanquan Gu}{quanquan.gu@bytedance.com}

\icmlkeywords{Machine Learning, ICML}

\vskip 0.3in
]



\printAffiliationsAndNotice{\icmlEqualContribution} 

\begin{abstract}
The conformational landscape of proteins is crucial to understanding their functionality in complex biological processes.
Traditional physics-based computational methods, such as molecular dynamics (MD) simulations, suffer from rare event sampling and long equilibration time problems, hindering their applications in general protein systems.
Recently, deep generative modeling techniques, especially diffusion models, have been employed to generate novel protein conformations. However, existing score-based diffusion methods cannot properly incorporate important physical prior knowledge to guide the generation process, causing large deviations in the sampled protein conformations from the equilibrium distribution. 
In this paper, to overcome these limitations, we propose a force-guided $\mathrm{SE}(3)$ diffusion model, \model, for protein conformation generation. 
By incorporating a force-guided network with a mixture of data-based score models, \model can generate protein conformations with rich diversity while preserving high fidelity.
Experiments on a variety of protein conformation prediction tasks, including 12 fast-folding proteins and the Bovine Pancreatic Trypsin Inhibitor (BPTI), demonstrate that our method surpasses the state-of-the-art method.
\end{abstract}

\section{Introduction}

Proteins are dynamic macromolecules that play pivotal roles in various biological processes. Their functionality is realized primarily through conformational changes--structural alterations that enable proteins to interact with other molecules. Depicting the protein conformational landscape provides vital insights for (1) identifying potential druggable sites hidden beneath the protein surface, and (2) revealing transition pathways between multiple metastable states. A comprehensive understanding of protein conformations facilitates the elucidation of biological reaction mechanisms, thereby empowering researchers to design targeted inhibitors and therapeutic agents with improved specificity and efficacy.

Traditional physics-based simulation methods such as molecular dynamics (MD) simulations have been extensively studied for protein conformation sampling. With a well-designed empirical force field and numerical integrator, the model propagates the 3D structure of a protein system over time following Newtonian mechanics. MD simulations converge towards the equilibrium distribution (i.e., the Boltzmann distribution) given sufficient time, which facilitates estimation of significant thermodynamic properties, e.g., binding free energy change \citep{fep}.
However, to preserve energy conservation and ensure numerical stability, the time step of MD simulations is typically only a few femtoseconds. This poses a challenge as certain biological processes of interest, such as protein folding, span much longer timescales, ranging from microseconds to seconds \citep{anton3}. This results in limited sampling efficiency within conventional MD simulations, further compounded by the rare event sampling problem \citep{rare_event_sampling}, impeding the research community to widely adopt MD for high throughput studies.


Building upon the cornerstone of powerful folding models (e.g., AlphaFold \citep{alphafold}, RoseTTAFold \citep{RoseTTAFold}, OmegaFold \citep{omegafold}, etc.), several attempts have been made to tailor these deep neural networks for protein conformation sampling. By perturbing the model input, such as multiple sequence alignment (MSA) masking \citep{speachaf} or clustering \citep{afcluster}, the folding model provides a more diverse set of possible folded structures, i.e., alternative conformations. However, this heuristic approach cannot guarantee the predicted structure to be a low energy state of the target sequence.
More recently, several works have incorporated diffusion models \citep{diffusion1, diffusion2} for protein conformation generation \citep{Eigenfold,str2str,DiG}. By pretraining on a large amount of known protein structures and efficient sampling through a predefined stochastic process, these models have shown promise in exploring diverse protein conformational states.
Nevertheless, existing diffusion models fall short in utilizing important physical prior information, such as the MD force field, to guide their diffusion process, hampering their capability to faithfully sample diverse protein conformations complying with the Boltzmann distribution.

To address the aforementioned challenges, we propose a novel force-guided diffusion model, \model, aiming to generate high fidelity protein conformations that better adhere to the Boltzmann distribution.
Drawing inspiration from the contrastive energy prediction (CEP) technique \citep{CEP}, we employ the MD energy prior as a \textit{physics-based preference function}. By introducing an additional force guidance network during the diffusion sampling process, it prioritizes generating conformations with lower potential energy, which effectively enhances sampling quality.
Our model is trained on general protein structures from the Protein Data Bank (PDB) \citep{pdb} as well as self-generated conformation samples, without relying on MD simulation data \citep{DiG}. To sum up, the main contributions of this work are highlighted as follows:

%
\begin{itemize}[leftmargin=*]
\setlength\itemsep{-0.5em}
\item Employing a sequence-conditional model to guide the unconditional model, we use classifier-free guidance on $\mathrm{SE}(3)$ to find a better trade-off between conformation quality and diversity. Compared with DiG \citep{DiG}, our method does not rely on MD data during training; compared with Str2Str \citep{str2str}, the guidance intensity coefficient provides a higher degree of freedom for balancing sample diversity and quality.


\item We utilize the MD energy function as a \textit{physics-based} reward to guide the generation of protein conformations. In addition, we propose an intermediate force guidance strategy during the diffusion sampling process. To the best of our knowledge, this is the first force-guided network suitable for protein conformation generation, contributing to the alignment of diverse conformation predictions with the equilibrium distribution.

\item Experiments on a variety of benchmarks demonstrate that our method outperforms the state-of-the-art approaches. In particular, energy and force guidance effectively guide the model to sample conformations with lower energy, leading to diverse samples more truthful to the underlying Boltzmann distribution.
\end{itemize}

\section{Related Work}

\noindent\textbf{Protein Conformation Prediction.} Perturbing pretrained folding models, such as AlphaFold \citep{alphafold}, to obtain a diverse set of alternative conformations marks the first attempt to use deep neural networks for multi-conformation predictions. \citet{speachaf} introduced mutations to the MSA representations to obtain different folded structures from AlphaFold. Similarly, \citet{afcluster} clusters MSA by sequence similarity to enable AlphaFold to discover alternative folding states of known metamorphic proteins. Reducing the MSA depth can also unlock multi-conformation prediction capability of AlphaFold \citep{del2022sampling}.
In addition, \citet{af2-rave} proposed to use the outputs from AlphaFold as initialization for AI-augmented MD simulations. \citet{bg}, \citet{idpgan} and \citet{ensvae} utilized MD simulation data to generate protein conformation ensembles.

Recently, diffusion models have been  employed for protein conformation generation. \citet{DiG} proposed Distributional Graphormer (DiG), which is trained on both protein structures from the PDB and MD simulation data. Unlike our proposed method, DiG incorporates an additional regularization 
into its loss function to align the learned score with MD force field at small diffusion time, which then extends over the entire pathway by the Fokker-Planck equation.
\citet{Eigenfold} introduced \textsc{EigenFold}, a harmonic diffusion model with physics-inspired prior to sample protein conformations. The cascading-resolution generative process allows efficient sampling across proteins of varying length. The model achieves remarkable performance in a number of benchmark tasks, yet the advantage of using a harmonic prior over the conventional isotropic Gaussian prior is not evidently clear.
Inspired by simulated annealing,  \citet{str2str} proposed Str2Str, a heating-annealing generative framework using an unconditional score model. By adjusting the duration $T_{\delta}$ of heating (i.e., forward diffusion) process, a certain degree of sample diversity could be achieved. However, during the model's training/inference phase, the absence of sequence or energy information evokes a question of whether the diverse generated outcomes adhere to the Boltzmann distribution. 

\noindent\textbf{Diffusion Models for Protein Design.} Another line of work focuses on developing diffusion models for protein design \citep{protdiff,anand2022protein,foldingdiff}. In particular, \citet{RFdiffusion} repurposed RoseTTAFold \citep{RoseTTAFold} to generate novel protein-binder backbones with successful experimental validation. \citet{chroma} proposed Chroma, which introduces a diffusion process that respects the conformational statistics of polymer ensembles, and can be effectively conditioned on protein semantics or even natural language to generate structures with desired properties. \citet{diffusionse3} proposed framediff, an innovative diffusion process on $\mathrm{SE}(3)$ for equivariant protein backbone generation, and has been recently extended to the flow-matching paradigm \citep{frameflow,bose2023se}. 
Diffusion models have also been applied to antibody sequence-structure co-design \citep{abdiffuser,diffab}.
The effectiveness of these approaches underscores the potential for further advancements of diffusion modeling in protein studies.

\noindent\textbf{Controllable Generation via Guided Diffusion.}
Controllable generative modeling is key to aligning diffusion models with human preference in many real-world tasks. Both classifier guidance \citep{classifierguidance,diffusion1} 
and classifier-free guidance \citep{freeguidance} have been proposed to guide an unconditional model with a preferred conditional variable, showing remarkable performance in a wide range of applications including text-to-image generation \citep{text-to-image,glide}, video generation \citep{imagen}, etc. Recently, \citep{CEP} proposed a novel energy guidance policy using a scalar reward function rather than fixed conditioning variables. By optimizing a contrastive energy prediction (CEP) objective, the model is guaranteed to converge to the exact guidance under enough model capacity and data samples. CEP shows great performance in image synthesis and reinforcement learning tasks. However, it is the gradient of energy function that is utilized as guidance in reverse sampling. This inspires us to propose a force guidance strategy which employs an equivariant network to directly approximate the intermediate force vector.

\section{Preliminaries}

\subsection{Protein Backbone Diffusion on SE(3)}\label{Diff-se3}

We adopt the protein backbone representation from AlphaFold \citep{alphafold}: for a protein with $N$ amino acid residues, its backbone atomic coordinates can be parameterized by a collection of $N$ orientation preserving rigid transformations (i.e., frames) to the local [N,C$_{\alpha}$,C,O] backbone atoms in each residue. We collectively denote the positions of all $N$ frames by
$\mathbf{x}_0 = [\mathbf{T}_0, \mathbf{R}_0] \in \mathrm{SE}(3)^N$, where $\mathbf{T}_0 \in \mathbb{R}^{3N}$  and $\mathbf{R}_0 \in \mathrm{SO(3)}^{N}$ denote the corresponding translation and rotation operations, respectively. With an additional backbone torsion angle $\psi$ describing the rotation of the oxygen atom around the C–C$_{\alpha}$ bond within each residue, we can reconstruct the protein backbone structure from the frame representations.

Following \citet{diffusionse3}, diffusion modeling on manifold $\mathrm{SE}(3)^N$ is employed for protein backbone generation. Two independent diffusion processes are defined for the translation and rotation subspaces, respectively:
\begin{align}
\mathrm{d} \mathbf{T}_t & = -\frac{1}{2} \beta_t \mathbf{P}\mathbf{T}_t \mathrm{~d} t+\sqrt{\beta_t} \mathbf{P} \mathrm{d} \mathbf{w}_t, \nonumber \\
\mathrm{~d}\mathbf{R}_t & = \sqrt{\frac{\mathrm{~d}}{\mathrm{~d}t} \sigma^2_t}\mathrm{~d} \mathbf{w}_t^{\mathrm{SO(3)}},  \label{eq:sde-r3-forward}
\end{align}
where subscript $t$ denotes the diffusion time variable in $[0, 1]$, $\beta_t$ and $\sigma_t$ are predefined time-dependent noise schedules, $\mathbf{P}$ is a projection operator removing the center of mass, and $[\mathbf{w}_t, \mathbf{w}_t^{\mathrm{SO(3)}}]$ represents the standard Wiener process in $[\mathcal{N}(0, I_3)^{\otimes N} , \mathcal{U}(\mathrm{SO}(3))^{\otimes N}]$.
The transition kernel of $\mathbf{T}$ satisfies $p_t(\mathbf{T}_t|\mathbf{T}_0)=\mathcal{N}(\mathbf{T}_t; \sqrt{\alpha_t} \mathbf{T}_0,(1-\alpha_t)\mathbf{I})$, where $\alpha_t = e^{-\int_0^t\beta_sds}$.
The rotational transition kernel satisfies $p_t(\mathbf{R}_t|\mathbf{R}_0)=\mathrm{IGSO}_3(\mathbf{R}_t;\mathbf{R}_0, t)$,
where $\mathrm{IGSO}_3$ is the isotropic Gaussian distribution on $\mathrm{SO}(3)$ \citep{igso3}.

The associated reverse-time stochastic differential equation (SDE) is as follows:
\begin{align}\label{eq:reverse-sde}
\mathrm{d} \mathbf{T}_t & = \mathbf{P}\left[-\frac{1}{2} \beta_t \mathbf{T}_t - \beta_t \nabla\log p_t(\mathbf{T}_t)\right]\mathrm{d}t +\sqrt{\beta_t}\mathbf{P} \mathrm{d}\bar{\mathbf{w}}_t, \nonumber \\
\mathrm{d} \mathbf{R}_t & = - \frac{\mathrm{~d}}{\mathrm{~d}t} \sigma^2_t \nabla\log p_t(\mathbf{R}_t) \mathrm{d} t+\sqrt{\frac{\mathrm{d}}{\mathrm{d}t} \sigma^2_t}\mathrm{d} \bar{\mathbf{w}}_t^{\mathrm{SO(3)}},
\end{align}
where $[\bar{\mathbf{w}}_t, \bar{\mathbf{w}}_t^{\mathrm{SO(3)}}]$ denotes another standard Wiener process in reverse time.

\subsection{Classifier-free Guidance for Diffusion Sampling}


Guided sampling has emerged as a critical strategy in developing diffusion models capable of generating samples complying with human instructions. Consider existing paired data $\mathbf{x}_0
\sim p_0(\mathbf{x}_0|c)$ with a conditioning variable $c$ (subscript denotes diffusion time, where $t=0$ corresponds to the original data), we typically perceive the conditional probability density at time $t$ as $p_t(\mathbf{x}_t|c)$. Applying Bayes’ rule we can obtain $p_t(\mathbf{x}_t|c) = \frac{p_t(\mathbf{x}_t)p_t(c|\mathbf{x}_t)}{p(c)}$. Therefore, one can train a classifier to predict the conditioning probability $p_t(c|\mathbf{x}_t)$ with given noisy data $\mathbf{x}_t$, and use the score of the classifier output as guidance \citep{classifierguidance}.
Instead of training a separate classifier to estimate $\nabla\log p_t(c|\mathbf{x}_t)$, \citet{freeguidance} proposed to utilize an implicit classifier $p^\gamma_t(c|\xb_t) \propto \big(\frac{p_t(\xb_t|c) p(c)}{p_t(\xb_t)} \big)^\gamma$, which then leads to a linear combination of an unconditional score estimator $s_{\theta}(\mathbf{x}_t)$ and a conditional score estimator $s_{\theta}(\mathbf{x}_t, c)$ to jointly estimate the target score function:
\begin{align*}
    & \nabla_{\mathbf{x}_t}\log p_t(\mathbf{x}_t|c) \\
    = & \nabla_{\mathbf{x}_t}\log p_t(\mathbf{x}_t) + \gamma\nabla_{\mathbf{x}_t}\log p_t(c|\mathbf{x}_t) \\
    = & \gamma\nabla_{\mathbf{x}_t}\log p_t(\mathbf{x}_t|c) + (1-\gamma) \nabla_{\mathbf{x}_t}\log p_t(\mathbf{x}_t) \\
    \approx & \gamma s_{\theta}(\mathbf{x}_t, c) + (1-\gamma) s_{\theta}(\mathbf{x}_t),
\end{align*}

where $\gamma$ is a hyperparameter controlling the guidance strength.  When $\gamma=0$, it reduces to an unconditional model, while at $\gamma=1$, it becomes a pure conditional model. These two models can be simultaneously trained under the same hood, where the unconditional model receives masked conditioning variable $c$ during training.

\section{Force-Guided Diffusion for Protein Conformation Generation}
\begin{figure}[t!]
    \centering
    \includegraphics[width=1.0\linewidth]{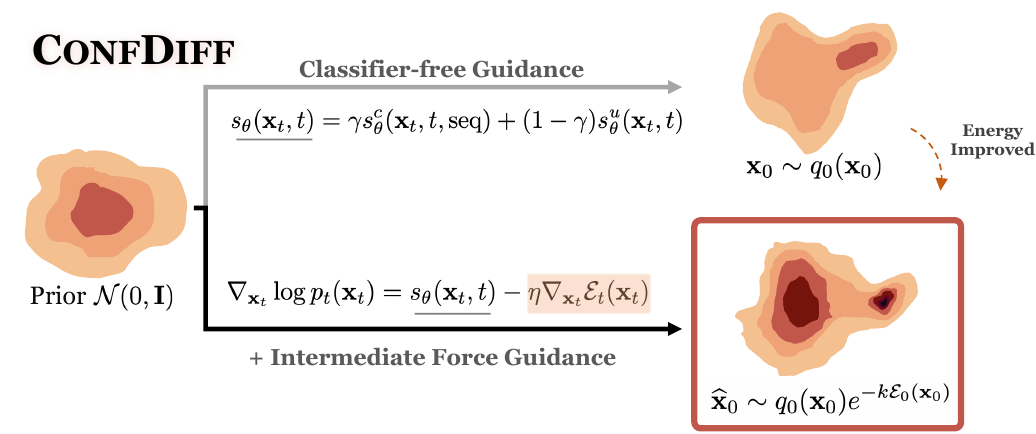}
    \caption{Protein conformation generation with multiple guidance strategies. \textbf{Upper:} With a mixture of sequence-conditional and unconditional score models, \model in Section~\ref{baseline} samples diverse conformations with reasonable quality. \textbf{Lower:} Incorporating force guidance in Section~\ref{force-guided}, the model generates structures with lower energy, better comply with the Boltzmann distribution.}
    \label{fig:overview}
\end{figure}

In this section, we propose force-guided \model, a diffusion model targeting multi-conformation generation for proteins. Employing a sequence-based conditional score network to guide an unconditional score model in Section~\ref{baseline}, \model achieves reasonable conformation diversity while ensuring sample quality. Building upon the energy guidance foundations in Section~\ref{energy-guided}, a novel force-guided sampling strategy is proposed to estimate the intermediate force function, which is then embedded within reverse time sampling process in Section~\ref{force-guided}. Utilizing prior information from the MD force field, our model successfully reweights the generated conformations to ensure they adhere better to the equilibrium distribution. A visual depiction is shown in Figure~\ref{fig:overview}.

\subsection{Sequence-Conditional Diffusion on $\mathrm{SE}(3)$}\label{baseline}

Our baseline model consists of an unconditional score model $s_{\theta}^{u}(\mathbf{x}_t, t)$ and a sequence-conditional one $s_{\theta}^{c}(\mathbf{x}_t, t, \mathrm{seq})$. 
The unconditional model is trained on protein structures (from the PDB) without any sequence information, effectively capturing the conformation distribution of general proteins.
On the other hand, the sequence-conditional model has access to both protein sequence information ($\mathrm{seq}$) and the corresponding structure $\mathbf{x}_t$ at time $t$. 

We adopt a similar network architecture to $\mathrm{FramePred}$ \citep{diffusionse3} to parameterize the corresponding score functions. The unconditional model takes sinusoidal embedding of the residue index and diffusion time $t$ as its single (\{$\mathbf{s}_i$\}) and pair (\{$\mathbf{z}_{ij}$\}) embeddings, where the conditional model additionally concatenates precomputed representations from ESMFold \citep{esmfold} to its single embedding. Note that the choice of sequence representation for the conditional model is flexible – it has been shown that using pretrained representations from folding models helps diffusion models generate reasonable protein structures \citep{Eigenfold,DiG}, 
while the unconditional model can effectively improve sampling diversity.
Both models have been trained with the denoising score matching (DSM) loss function:
\begin{align*}
\mathcal{L}_{\mathrm{DSM}} = \mathbb{E}\left[\lambda(t) \| s_{\theta}(\mathbf{x}_t, t) - \nabla_{\mathbf{x}_t}\log p_t(\mathbf{x}_t|\mathbf{x}_0)\|^2_2 \right],
\end{align*}
where $\lambda(t)$ is a reweighting function inversely proportional to the score norm \citep{diffusion1}.
A detailed illustration of our model architecture is provided in Appendix \ref{appendix:sde}.

During the reverse sampling process, we use a hyperparameter $\gamma$ to control the classifier-free guidance strength from the conditional model, so that the score function $\nabla_{\mathbf{x}_t}\log p_t(\mathbf{x}_t|\mathrm{seq})$ is estimate by
\[
s_{\theta}(\mathbf{x}_t,t|\mathrm{seq}) = \gamma s^c_{\theta}(\mathbf{x}_t, t, \mathrm{seq}) + (1-\gamma) s^u_{\theta}(\mathbf{x}_t, t).
\]
For notation simplicity, we hereafter omit the sequence conditional term in our baseline score model, i.e., $s_{\theta}(\mathbf{x}_t, t) = s_{\theta}(\mathbf{x}_t, t|\mathrm{seq})$.


\subsection{Intermediate Energy-Guided Diffusion} \label{energy-guided}

Despite the diffusion model's capability to generate diverse structures, these conformations are not always reasonable in the sense that they may reside in the high energy region of the potential energy surface.
Due to the limited availability of multi-conformation data within existing protein structure databases, the training data distribution does not comply with the equilibrium distribution, but rather only containing a few data points near the potential energy minima for each protein sequence.
This necessitates the development of a generative model propelled by data and steered by physics-based guidance towards generating samples according to the Boltzmann distribution.
To that end, we introduce energy guidance, where the MD energy function attributes a reward to guide the conformation generation process.

Given an existing (baseline) diffusion model which generates samples $\mathbf{x}_0 \sim q_0(\mathbf{x}_0)$, our goal is to sample protein conformations from the equilibrium distribution
\begin{align}\label{eq:reweight}
    p_0(\mathbf{x}_0) = q_0(\mathbf{x}_0)\frac{e^{-k\mathcal{E}_0 (\mathbf{x}_0)}}{Z},
\end{align}
where $Z:=\int q_0(\mathbf{x}_0)e^{-k\mathcal{E}_0 (\mathbf{x}_0)}\mathrm{~d}\mathbf{x}_0$ is the intractable normalizing constant, $\mathcal{E}_0$ is the \texttt{OpenMM} \citep{eastman2017openmm} energy function which evaluates the potential energy of each generated conformation $\mathbf{x}_0$. 
$k$ is the inverse temperature factor.
Given any test function $F(\mathbf{x}_0)$, $p_0(\mathbf{x}_0)$ gives a more accurate estimate through importance sampling, defined as $\mathbb{E}_{p_0(\mathbf{x}_0)}[\frac{F(\mathbf{x}_0)}{e^{-k\mathcal{E}_0(\mathbf{x}_0)}/Z}]$.



In the forward diffusion process, we perturb the original samples $\mathbf{x}_0\sim q_0(\mathbf{x}_0)$ according to the SDE described in Eq.~(\ref{eq:sde-r3-forward}), and we assert that $p_t(\mathbf{x}_t|\mathbf{x}_0) := q_t(\mathbf{x}_t|\mathbf{x}_0)$. Whereas, it is the marginal distribution $p_t(\mathbf{x}_t)$ that truly captures our attention. According to \citet{CEP}, we have the following properties for the energy function:

\begin{proposition} \label{prop1}
Suppose $p_0(\mathbf{x}_0) = q_0(\mathbf{x}_0)\frac{e^{-k\mathcal{E}_0 (\mathbf{x}_0)}}{Z}$, and for $t\in (0, 1]$, $p_t(\mathbf{x}_t|\mathbf{x}_0):=q_t(\mathbf{x}_t|\mathbf{x}_0)$. Then, the marginal distribution satisfies $p_t(\mathbf{x}_t)\propto q_t(\mathbf{x}_t) e^{-k \mathcal{E}_t(\mathbf{x}_t)}$, where $q_t(\mathbf{x}_t)$ is the data-based marginal distribution, and $\mathcal{E}_t(\mathbf{x}_t)$ satisfies
\begin{align}\label{eq:energy_t}
    \mathcal{E}_t\big(\mathbf{x}_t\big)= -\frac{1}{k}\log \mathbb{E}_{q_{t} \left(\mathbf{x}_0|\mathbf{x}_t\right)}\left[e^{-k \mathcal{E}_0\left(\mathbf{x}_0\right)}\right].
\end{align}
\end{proposition}
We refer to $\mathcal{E}_t$ as the intermediate energy function, capturing the dynamic changes in energy at various stages during the diffusion process.

\par With the exact formula in Eq.~(\ref{eq:energy_t}), a neural network $f_{\phi}(\mathbf{x}_t, t)$
parameterized by $\phi$ is utilized to approximate the intermediate energy function $\mathcal{E}_t(\mathbf{x}_t)$ by the optimization problem in \citet{CEP}
\begin{small}
\begin{align*}
\mathcal{L}_{\mathrm{CEP}} = \mathbb{E}_{p(t)}\mathbb{E}_{q_0(\mathbf{x}_0), q_t(\mathbf{x}_t|\mathbf{x}_0)}\left[-e^{-k \mathcal{E}_0(\mathbf{x}_0)}\log\frac{e^{-f_{\phi}(\mathbf{x}_t, t)}}{\sum e^{-f_{\phi}(\mathbf{x}_t, t)}} \right].
\end{align*}
\end{small}

It has been proved in \citet{CEP} that the optimal $f_{\phi^*}(\mathbf{x}_t, t)$ satisfies $\nabla_{\mathbf{x}_t}f_{\phi^*}(\mathbf{x}_t, t) = k \nabla_{\mathbf{x}_t}\mathcal{E}_t(\mathbf{x}_t)$, so that sampling can be carried out according to the reverse-time SDE with score function:
\begin{eqnarray*}
\nabla_{\mathbf{x}_t}\log p_t(\mathbf{x}_t) = s_{\theta}(\mathbf{x}_t, t) - \eta\nabla_{\mathbf{x}_t} f_{\phi}(\mathbf{x}_t, t),
\end{eqnarray*}
where $\eta$ is the hyperparameter controlling the guidance strength, and $s_{\theta}(\mathbf{x}_t, t)$ refers to the baseline score model in Section~\ref{baseline}. Note that $s_{\theta}(\mathbf{x}_t, t)$ and the intermediate energy network $f_{\phi}(\mathbf{x}_t, t)$ are trained independently.

Since our score network only samples protein backbone structures, we use \texttt{faspr} \cite{huang2020faspr} to generate full-atom structures for energy evaluation using \texttt{OpenMM} \cite{eastman2017openmm}. See implementation details in Appendix \ref{mtd:openmm}. However, we come across inherent challenges with normalization intricacies in the energy computed by \texttt{OpenMM}. Within limited batch size, we witness significant numerical volatility in estimating. In contrast, the numerical value of atomic force will be much more stable. 
%
%
Consequently, the design of a force-guided policy tailored for protein conformation generation proves to be incredibly significant. Compared with energy guidance, force guidance does not suffer from the energy fluctuation problem and can be directly applied as guidance in the reverse sampling process.
In the next section, we will illustrate how to realize intermediate force guidance.


\subsection{Intermediate Force-Guided Diffusion} \label{force-guided}

In the context of protein conformation modeling, we have access to both a physics-based energy function $\mathcal{E}_0 (\mathbf{x}_0)$ (i.e., interatomic potential energy) as well as its gradient  $\nabla_{\mathbf{x}_0}\mathcal{E}_0(\mathbf{x}_0)$ (i.e., force on each atom). In contrast to the unnormalized potential energy function, atomic force is more local and exhibits better numerical stability, which also aligns better with the score matching objective. Following the marginal distribution of energy in Eq.~\eqref{eq:energy_t}, we have the following properties for the force function.


\begin{proposition} \label{prop2}
Given the presumption of $p_t(\mathbf{x}_t|\mathbf{x}_0) := q_t(\mathbf{x}_t|\mathbf{x}_0)$, $0<t\le 1$, the energy function of intermediate state follows Eq.~\eqref{eq:energy_t}. The corresponding intermediate force has the following formula
\begin{eqnarray}\label{eq:force_t}
 \nabla_{\mathbf{x}_t}\mathcal{E}_t\big(\mathbf{x}_t\big)= \frac{\mathbb{E}_{q_t(\mathbf{x}_0|\mathbf{x}_t)}\left[ e^{-k \mathcal{E}_0(\mathbf{x}_0)} \zeta(\mathbf{x}_0,\mathbf{x}_t)\right]}{k\mathbb{E}_{q_t(\mathbf{x}_0|\mathbf{x}_t)}\left[e^{-k \mathcal{E}_0(\mathbf{x}_0)} \right]},
\end{eqnarray}
where $\zeta(\mathbf{x}_0, \mathbf{x}_t)=\nabla_{\mathbf{x}_t}\log q_t(\mathbf{x}_t)-\nabla_{\mathbf{x}_t}\log q_t(\mathbf{x}_t|\mathbf{x}_0)$. Denoting implicit distribution $p_t(\mathbf{x}_t)\propto q_t(\mathbf{x}_t) e^{-\eta k\mathcal{E}_t(\mathbf{x}_t)}$, the score function follows
\begin{eqnarray}\label{eq:score_guidance}
    \nabla_{\mathbf{x}_t}\log p_t(\mathbf{x}_t) =\nabla_{\mathbf{x}_t}\log q_t(\mathbf{x}_t) - \eta\nabla_{\mathbf{x}_t}\mathcal{E}_t(\mathbf{x}_t),
\end{eqnarray}
where $\eta$ is a hyperparameter controlling the strength of force guidance.

\end{proposition}
Detailed derivations are provided in Appendix \ref{derivation}.

Proposition \ref{prop2} unveils the precise force at time $t$, thereby advancing our understanding of the intermediate energy function.
The intermediate force formula consists of the ground truth potential energy, the diffusion transition kernel, and a marginal score function (estimated by the score model in Section~\ref{baseline}). As $t$ approaches $0$, $q_t(\mathbf{x}_0|\mathbf{x}_t)$ converges to the delta function and the intermediate force $\nabla_{\mathbf{x}_t} \mathcal{E}_t(\mathbf{x}_t)$ converges to the ground truth force. On the other hand, when $t$ approaches $1$, $q_t(\mathbf{x}_t|\mathbf{x}_0)$ reduces to $q_t(\mathbf{x}_t)$, and the force $\nabla_{\mathbf{x}_1} \mathcal{E}_1(\mathbf{x}_1)$ vanishes to $0$. A comprehensive illustration is provided in Appendix \ref{derivation}.

\begin{algorithm}[t]
    \caption{Force-guided \model}
    \label{alg:force}
    \textbf{Input:} generated data $\mathbf{x}_0$ in a batch $K$, score model $s_{\theta}(\mathbf{x}_t, t)$ in Section \ref{baseline}, energy $\mathcal{E}_0(\mathbf{x}_0)$,  force $\nabla_{\mathbf{x}_0}\mathcal{E}_0(\mathbf{x}_0)$, intermediate force network $h_{\psi}(\mathbf{x}_t, t)$
    \begin{algorithmic}[1]
    \FOR{training iterations}
        \STATE $t\sim \mathcal{U}(0,1)$.
        \STATE $\mathbf{x}_t =\mathrm{Forward\ diffusion\ in}$ Eq.~(\ref{eq:sde-r3-forward}). 
        \STATE $q_t(\mathbf{x}_t|\mathbf{x}_0)\sim \mathcal{N}(\mathbf{x}_t; \sqrt{\alpha_t}\mathbf{x}_0, (1-\alpha_t)I)$ 
        \STATE $\zeta(\mathbf{x}_0, \mathbf{x}_t)= s_{\theta}(\mathbf{x}_t, t) - \nabla\log q_t(\mathbf{x}_t|\mathbf{x}_0)$.
        \STATE $Y=\sum_{\mathbf{x}_0} q_t(\mathbf{x}_t|\mathbf{x}_0)e^{-k\mathcal{E}_0(\mathbf{x}_0)}$.
        \STATE $\mathcal{L} = \frac{1}{K}\sum \|h_{\psi}(\mathbf{x}_t, t) - {e^{-k\mathcal{E}_0(\mathbf{x}_0)}\zeta(\mathbf{x}_0, \mathbf{x}_t)}/Y\|^2_2$. 
        \STATE $\min_{\psi} \mathcal{L}$.
    \ENDFOR
    \end{algorithmic}
\end{algorithm}

\paragraph{Training process.} We employ our baseline score model introduced in Section~\ref{baseline} to generate protein conformations from $q_0(\mathbf{x}_0)$, which are then used to train an independent intermediate force network $h_{\psi}(\mathbf{x}_t, t): \mathcal{M} \times (0,1] \rightarrow \mathrm{Tan}_{\mathbf{x}_t}\mathcal{M}$, the tangent space of $\mathcal{M}$ at $\mathbf{x}_t$. 


Let $K$ be a positive integer denoting the training batch size. To ensure that data in a batch adheres to the same Boltzmann distribution, we first randomly choose a protein sequence, then subsequently draw $K$ samples from this sequence. By adding noise to the sampled data $\mathbf{x}_0$ following the SDE in Eq.~(\ref{eq:sde-r3-forward}), we obtain perturbed data at time $t$ as $\mathbf{x}_t = \sqrt{\alpha_t}\mathbf{x}_0 + \sqrt{1-\alpha_t}\mathbf{\epsilon}_t, \mathbf{\epsilon}_t\sim\mathcal{N}(0, I)$, such that $q_t(\mathbf{x}_t|\mathbf{x}_0)\propto 1/\sqrt{1-\alpha_t}\exp\left(-\|\mathbf{x}_t - \sqrt{\alpha_t}\mathbf{x}_0\|^2 / 2\right)$. 
We define the intermediate force loss function $\mathcal{L}_{\mathrm{f}}(\psi)$ as
\begin{align}\label{eq:loss_force}
    \mathbb{E}_{p(t),\mathbf{x}_0, \mathbf{x}_t}\bigg[ \bigg\| h_{\psi}(\mathbf{x}_t, t) - \frac{e^{-k \mathcal{E}_0(\mathbf{x}_0)}\zeta(\mathbf{x}_0, \mathbf{x}_t)}{\mathbb{E}_{\mathbf{x}_0^K}\left[q_t(\mathbf{x}_t|\mathbf{x}_0) e^{-k \mathcal{E}_0(\mathbf{x}_0)}\right]} \bigg\|^2_2 \bigg],
\end{align}
where $\zeta(\mathbf{x}_0, \mathbf{x}_t)=\nabla_{\mathbf{x}_t}\log q_t(\mathbf{x}_t) - \nabla_{\mathbf{x}_t}\log q_t(\mathbf{x}_t|\mathbf{x}_0)$. The latter component of Eq.~(\ref{eq:loss_force}) signifies the precise value of the intermediate force at time $t$, where $\nabla_{\mathbf{x}_t}\log q_t(\mathbf{x}_t)$ is the intractable score function estimated by $s_{\theta}(\mathbf{x}_t, t)$ in our baseline and $q_t(\mathbf{x}_t|\mathbf{x}_0)$ is the tractable Gaussian distribution. Instead of computing $\mathbb{E}_{\mathbf{x}_0}\left[ q_t(\mathbf{x}_t|\mathbf{x}_0) e^{-\mathcal{E}_0(\mathbf{x}_0)}\right]$ in the entire generated data collection, we estimate the expectation with the $K$ samples in a batch \citep{CEP}.
The force-guided training process is summarized in Algorithm~\ref{alg:force}.


Taking into account that the exact intermediate force formula in Eq.~(\ref{eq:force_t}) equals to the MD force-field $\nabla_{\mathbf{x}_0} \mathcal{E}_0(\mathbf{x}_0)$ at $t=0$, and turns $0$ when $t=1$, we construct the network $h_{\psi}(\mathbf{x}_t, t)$ as the interpolation form following \citep{mate2023learning}:
\begin{align*}
    h_{\psi}(\mathbf{x}_t, t) = (1-t)\nabla_{\mathbf{x}_0} \mathcal{E}_0(\mathbf{x}_0) + t(1-t) g_{\psi}(\mathbf{x}_t, t),
\end{align*}
where $g_{\psi}(\mathbf{x}_t, t)$ is a neural network estimating the intermediate term within our interpolation construction. It ensures the boundary conditions at $t\in\{0, 1\}$ and we empirically find that this construction reduces the variance during training. Detailed illustration is shown in Appendix~\ref{appendix:interpolation}.

\paragraph{Inference process.} Starting from the prior distribution in $\mathrm{SE}(3)^N$, the entire inference process is illustrated in Algorithm~\ref{alg:inference}. Score network $[S_{\mathbf{T}_i}, S_{\mathbf{R}_i}]$ is the baseline model mixing the sequence conditional score and unconditional score, as introduced in Section~\ref{baseline}. Force guidance term $h_{\psi}(\mathbf{x}_t, t)$ is incorporated to the score function term at every time step during the reverse sampling process, whose guidance strength is controlled by hyperparameter $\eta$.
To maintain structural stability, force guidance is only applied to the translational components (i.e., $\alpha$-carbons) in $\mathbb{R}^{3N}$.

\begin{algorithm}[t]
    \caption{Inference with force-guided \model}
    \small
    \label{alg:inference}
    \begin{algorithmic}[1]
    \STATE $\mathrm{x}_N=[\mathrm{T}_N, \mathrm{R}_N]\sim [\mathcal{N}(0, I_3)^{\otimes N} , \mathcal{T}\mathcal{N}_{\mathbf{R}}(0, I)^{\otimes N}]$. \\
    \# $\mathcal{T}$: the tangent space
    \FOR{SDE time step $i=N,...,1$}
        \STATE $\epsilon\sim [\mathcal{N}(0, I_3)^{\otimes N} , \mathcal{T}\mathcal{N}_{\mathbf{R}}(0, I)^{\otimes N}]$. \\
        \STATE $[S_{\mathbf{T}_i}, S_{\mathbf{R}_i}]=\gamma s_{\theta}(\mathbf{x}_i, i, \mathrm{seq}) + (1-\gamma) s_{\theta}(\mathbf{x}_i, i)$. 
        \STATE $H=h_{\psi}(\mathbf{x}_i, i)$.
        \STATE $M_{\mathbf{T}_i}= \mathbf{P}(\frac{1}{2}\beta_i\mathbf{T}_i + \beta_i(S_{\mathbf{T}_i}-\eta H)+\sqrt{\beta_i}\epsilon_{\mathbb{R}^{3N}})$.
        \STATE $M_{\mathbf{R}_i} = \frac{\mathrm{~d}}{\mathrm{~d}i} \sigma^2_i \cdot S_{\mathbf{R}_i} + \sqrt{\frac{\mathrm{~d}}{\mathrm{~d}i} \sigma^2_i}\epsilon_{\mathbb{SO}(3)^N}$.
        \STATE $\mathbf{x}_{i-1} = \exp_{\mathbf{x}_i}\{[M_{\mathbf{T}_i}, M_{\mathbf{R}_i}]\}$.
    \ENDFOR
    \STATE $\mathrm{x}_0=[\mathrm{T}_0, \mathrm{R}_0]$.
    \end{algorithmic}
\end{algorithm}

\vspace{0.5em}
\section{Experiments}

We evaluate the effectiveness of \model with different guidance strategies on protein conformation generation tasks.
In brief, we demonstrate that 1) classifier-free guidance balances the trade-off between sample quality and diversity; 2) energy and force guidance effectively guide the model to sample conformations with lower energy; 3) combining classifier-free and physical guidance strategies, \model outperforms current state-of-the-art models by generating high-quality and diverse samples more truthful to the underlying Boltzmann distribution. Code and checkpoints are open-sourced at \url{https://github.com/bytedance/ConfDiff}.



\subsection{Experimental Setup}

\noindent\textbf{Models.} We compare \model with two state-of-the-art diffusion-based models for protein conformation generation: \textsc{EigenFold} \citep{Eigenfold} samples protein conformations using harmonic diffusion and is fully conditioned on pretrained sequence representations; \textsc{Str2Str} \citep{str2str} first predicts a folded structure using a protein folding model, then perturbs the structure through forward-and-backward diffusion of an unconditional model (sequence-agnostic). All baseline models are run on their default settings. For \model, we follow \textsc{Str2Str} to sample conformations with varying sequencing condition levels ($\gamma$). All models are trained on experimental protein structures deposited in the Protein Data Bank (PDB) without using additional MD trajectory data.

\vspace{0.3em}

\noindent\textbf{Datasets.} 
We apply \model on two protein conformation benchmarks for a comprehensive evaluation: 1) \textbf{fast-folding proteins} comprise MD simulation data for 12 small proteins with fast folding-unfolding dynamics \cite{deshawfastfold2011}, 2) bovine pancreatic trypsin inhibitor (\textbf{BPTI}) contains MD simulation data for BPTI exhibiting five metastable states \citep{deshawbpti2010}. See Appendix \ref{si:datasets} for more details.


\begin{table*}[bt]
\caption{Results on fast-folding proteins. The results for \textit{classifier-free}, \textit{energy guidance}, and \textit{force guidance} are labeled as \textsc{\model-Base}, \textsc{\model-Force}, and \textsc{\model-Energy}. All values are shown as mean/median. Best result for each metric is shown in \textbf{bold} and the second best is \underline{underlined}.}
\label{tab:fastfold-main}
\vskip 0.15in
\begin{center}
\begin{small}
\begin{sc}
\begin{tabular}{lccccccc}
\toprule
\multirow{2}{*}{Models}         & \multicolumn{4}{c}{JS distance ($\downarrow$)}                      &  Val-CA         &  RMSE$_\text{contact}$    & RMSF   \\ \cmidrule(lr){2-5} 
                & PwD            &  Rg                  & TIC         & TIC-2D        &  ($\uparrow$)    &  ($\downarrow$)      &   (\si{\angstrom})  \\
\midrule
EigenFold   & 0.53/0.56 & 0.52/0.55 & 0.50/0.50 & 0.64/0.66 & 0.15/0.08 & 6.18/6.22 & 1.6/1.1 \\
Str2Str-SDE & 0.34/0.32 & 0.30/0.24 & 0.39/0.38 & 0.56/0.58 & \textbf{0.97/0.98} & 3.68/4.01 & 7.8/8.0 \\
Str2Str-ODE & 0.37/0.38 & 0.33/0.30 & 0.40/0.39 & 0.57/0.59 & 0.96/0.97 & 4.14/4.36 & 6.4/6.3 \\
\midrule
\rowcolor{Highlight}
\model-Base   & \textbf{0.29/0.27} & \textbf{0.25/0.22} & \textbf{0.36/0.37} & \textbf{0.52/0.52} & 0.89/0.91 & \underline{3.61/3.57} & 6.1/5.9 \\
\rowcolor{Highlight}
\model-Energy  & 0.34/0.34 & 0.31/0.29 & 0.39/0.40 & 0.54/0.56 & 0.97/0.97 & 3.65/3.80 & 7.1/6.1   \\
\rowcolor{Highlight}
\model-Force  & \textbf{0.29/0.27} & \underline{0.26/0.24} & \underline{0.38/0.38} & \underline{0.54/0.54} & \textbf{0.97/0.98} & \textbf{3.25/3.38} & 6.2/5.7 \\
\bottomrule
\end{tabular}
\end{sc}
\end{small}
\end{center}
\end{table*}

\vspace{0.5em}
\noindent \textbf{Metrics.}
We evaluate generated conformations as to their \textit{validity}, \textit{precision}, \textit{diversity} and the \textit{similarity of distributions} when comparing conformation ensembles. We summarize key metrics here and include details in Appendix \ref{si:metrics}:
\begin{itemize}[leftmargin=*]
    \setlength\itemsep{-0em}
    \item \textit{Validity} evaluates whether generated conformations comply with basic physics as valid proteins. Similar to \citet{str2str}, we evaluate if a conformation contains steric \textit{clashes} between any residues or \textit{bond breaks} between adjacent residues, based on the coordinates of $\alpha$-carbons (CA). Conformations without \textit{clash} and \textit{break} are considered valid structures and we calculate the validity \textsc{Val-CA} as the fraction of valid conformations in all generated samples. 
    \item \textit{Precision} indicates how accurate the model can recover known conformations of a protein. To evaluate structural similarity, sample conformations are aligned to a reference structure and the root-mean-square deviation (RMSD) of $\alpha$-carbons is calculated. The smallest RMSD between samples and the reference structure is reported as the \textit{precision}.
    \item \textit{Diversity} shows the model's ability to generate structurally different conformations. We estimate RMSD between pairs of conformations in samples and report the mean RMSD (termed \textsc{RMSF}) as sample diversity.
    \item \textit{Similarity between distributions} measures how well generated conformations resemble the samples from MD simulation (i.e., Boltzmann distribution). We follow \citet{str2str} to use Jensen-Shannon distance (JS) to characterize the similarity between generated samples and reference samples from MD. For distribution estimation, $\alpha$-carbon coordinates are projected to three types of feature dimensions: pairwise distance between $\alpha$-carbon atoms (\textsc{PwD}); radius-of-gyration (\textsc{Rg}) that is the distance of a $\alpha$-carbon atom to the center-of-mass, and time-lagged independent components (\textsc{TIC}), a reduced feature space related to protein dynamics. We compute JS distance on each feature dimension and report the mean distance; for \textsc{TIC}, we also report the JS distance between joint distributions projected to the first two TIC components (\textsc{TIC-2D}).
\end{itemize}

\vspace{-1.0em}
\subsection{Conformation Sampling with Force Guidance}
\begin{figure}[ht!]
    \centering
    \includegraphics[width=1.0\linewidth]{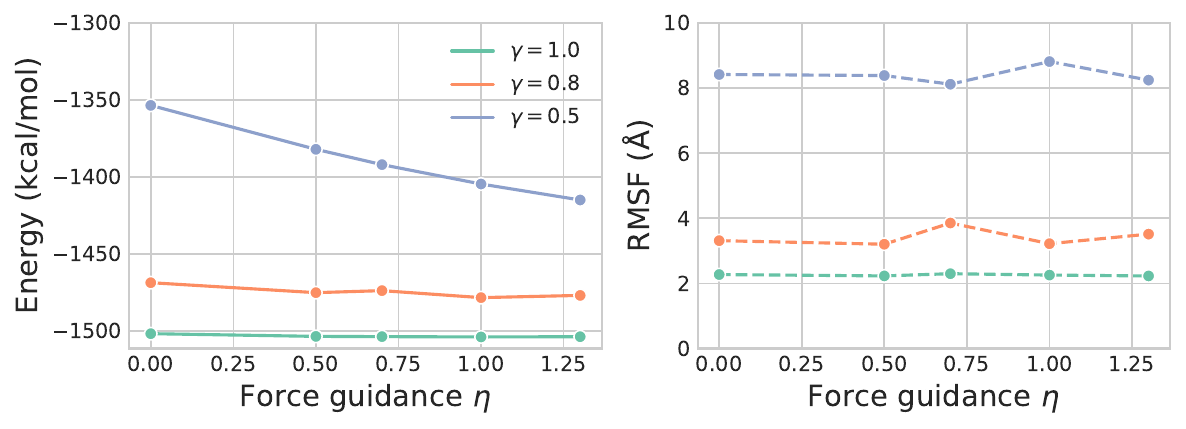}
    \caption{Energy (left) and diversity (right) of sampled conformations for WW-domain with various levels of force guidance ($\eta$) and sequence condition ($\gamma$). Models with weaker sequence condition generate more diverse samples and \textit{force guidance} improves conformation stability without drastically decease the diversity.} 
    \label{fig:energy-vs-eta}
\end{figure}

As an initial proof-of-concept, we first show that proposed \textit{force guidance} can improve the sampling capability of \model by generating protein conformations with reduced energy (i.e. improved stability). We select WW-domain, one of the fast-folding proteins from \citet{deshawfastfold2011} as an example and generate conformations using \model at varying degrees of force guidance ($\eta$) and sequence condition ($\gamma$). As shown in Figure~\ref{fig:energy-vs-eta}, without force guidance ($\eta=0$), models with weaker sequence condition can generate samples with higher diversity (RMSF), however, also with higher energy, showing the quality-diversity compromise. After integrating force guidance, \model mitigates such effect and are able to generates samples with lower energy while maintaining similar RMSF levels. This result shows that \textit{force guidance} can improve conformation stability without drastically decease the diversity. This favorable effect suggests integrating physical information might provide fine-grained guidance to sample locally optimized structures.
\vspace{0.5em}
\begin{figure}[hb!]
    \centering
    \includegraphics[width=1.0\linewidth]{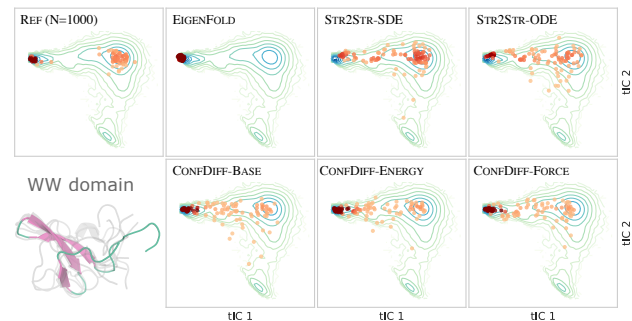}
    \caption{Sample distribution over the first two TIC components for WW-domain. \textsc{Ref (N=1000)} shows 1000 random samples from the reference MD simulation. The illustration in the lower left shows the experimental structure in the folded state (in color) and 5 random samples from the reference (in grey).}
    \label{fig:fastfold-main}
    \vskip 0.1in
\end{figure}





\begin{figure*}[ht!]
    \centering
\includegraphics[width=1.0\linewidth]{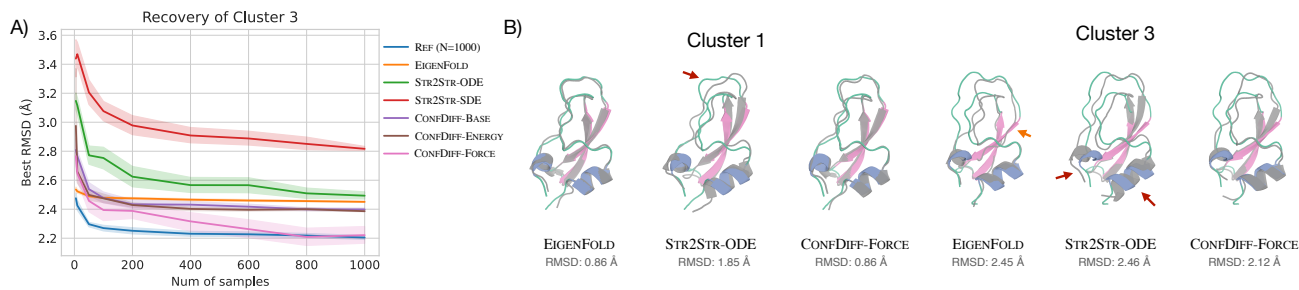}
    \caption{Metastable state prediction for BPTI. A) The precision of predicting Cluster 3 vs varying sample sizes. B) Visual comparison of best samples of three models for Cluster 1 and Cluster 3. Reference structures are shown in color and sample structures are in grey. RMSD vs reference is labeled. For Cluster 1, \textsc{Str2Str} shows lower accuracy in the upper loop region (red arrow); for Cluster 3, the more challenging task, \textsc{EigenFold} does not correctly predict the $\beta$-sheet (orange arrow) and \textsc{Str2Str} shows more global structural misalignment (red arrows) that might contribute to higher RMSD. }
    \label{fig:bpti-main}
\end{figure*}

\subsection{Distribution Prediction for Fast-Folding Proteins}
\vspace{-0.3em}
This dataset consists of 12 short proteins whose folding and unfolding dynamics have been examined in long-time all-atom MD simulations \citep{deshawfastfold2011}. In this experiment, we evaluate the capability of models to recover the conformation distributions observed in the simulation. This benchmark requires the model to not only predict the stable folded structures, but also physically viable unfolded or partially unfolded structures. To assess distribution similarity, we evaluate JS-distances between the distributions of generated samples and ground truth MD samples, along with the \textit{validity} score and \textsc{RMSF} for \textit{diversity}. In addition, we report the root-mean-square-error on the predicted contact rate between residues \textsc{RMSE$_\mathrm{contact}$} to reflect the model accuracy in flexible regions \cite{idpgan}.

We summarize results in Table \ref{tab:fastfold-main} and showcase the sample distribution in TIC projection in Figure~\ref{fig:fastfold-main} and Figure~\ref{fig:fastfold-tica-proj-all}. \model consistently outperforms \textsc{EigenFold} and \textsc{Str2Str} in terms of recovering sample distributions with lower JS distance and lower RMSE in predicting residue-residue contacts. Notably, integrating energy and force guidance improves the sample validity over \model-Base and sustains similar sample diversity, confirming the benefits of integrating physical guidance. 
~\\
In comparison, \textsc{EigenFold} shows limited sample diversity likely due to the strong tendency to predict the folded state as a fully-conditional diffusion model. \textsc{Str2Str} leverages the unconditional diffusion model to explore diverse structural space and generate diverse samples. Nevertheless, better distribution-related scores of \model suggests diffusion process controlled by sequence condition might be generate samples more truthful to a protein's true distribution.


\subsection{Metastable States Prediction for BPTI}
Previous MD simulation studies on BPTI have recovered five kinetic clusters, representing metastable states near its native folded state \citep{deshawbpti2010}. For this benchmark, we evaluate the quality and efficiency for models to recover these five states. Compared with fast-folding proteins, this benchmark requires model to generate different conformations while maintaining the correct folding structure.

We measure model \textit{precision} by the lowest sample RMSD to cluster centers. For comparison, we report the best RMSD averaged over five clusters (\textsc{RMSDavg}) and the RMSD to cluster 3 (\textsc{RMSDcls3}), the cluster that is the most difficult to sample.
As shown in Table~\ref{tab:bpti-main}, \model with force guidance performs the best on both metrics, suggesting better capability on predicting distinct metastable states.

\begin{table}[h]
\caption{Recovery of five metastable states for BPTI. \textsc{RMSDavg} denotes the best RMSD score, averaged over five states. \textsc{RMSDcls3} refers to the best RMSD to the remote cluster 3, which is the most difficult to sample. (Unit: \si{\angstrom}.)}
\label{tab:bpti-main}
\vskip 0.1in
\begin{center}
\begin{small}
\begin{sc}
\begin{tabular}{lcc}
\toprule
 Model   & RMSDavg ($\downarrow$) & RMSDcls3 ($\downarrow$)\\
\midrule
Reference & 1.10 & 2.19 \\
\midrule
EigenFold & 1.47 & 2.45 \\
Str2Str-ODE & 2.06 & 2.46 \\
Str2Str-SDE & 2.26 & 2.80 \\
\midrule
\rowcolor{Highlight}
\model-Base & 1.55 & 2.38 \\
\rowcolor{Highlight}
\model-Energy & 1.57 & 2.38  \\
\rowcolor{Highlight}
\model-Force & \textbf{1.41} & \textbf{2.12} \\
\bottomrule
\end{tabular}
\end{sc}
\end{small}
\end{center}
\end{table}

We further assess the model efficiency by comparing the precision at varying sample sizes. As shown in Figure~\ref{fig:bpti-main}A, all \model models show good efficiency to sample the remote cluster 3 while \textsc{EigenFold} quickly plateaued due to its low sample diversity. \model-Force showed best performance, and even close to \textsc{Ref (N=1000)}, 1000 conformations randomly sampled from MD trajectories.

These results show superb performance of \model models, particularly \textsc{\model-Force} in predicting distinct metastable states with high precision and potential to sample remote states.

\section{Conclusion and Future Work}
In this work, we introduce \model for protein conformation generation. By mixing a sequence-conditional score network with an unconditional model, we strike a balance between sampling quality and diversity through classifier-free guidance. Upon such a baseline model, we additionally propose novel physics-based energy and force guidance strategies with theoretical guarantee, which effectively guide the diffusion sampler to generate low energy conformations better complying with the underlying Boltzmann distribution. Experiments on a variety of protein systems demonstrate the effectiveness of our proposed method.

Although \model has shown promising performance, there is still room for further improvement.
Due to limited conformation diversity in existing protein structure databases, models fully conditioned on sequence information tend to predict the folded structure and lack the capability to fully picture the entire conformation landscape.
On the other hand, the full-atom MD energy function is still relatively computationally intensive and time-consuming to use as a preference function under the generative modeling framework.
Nevertheless, it is our first attempt to combine the Boltzmann prior with diffusion models and we will persist in the pursuit of more efficient approaches for force guidance, as well as effective measures to minimize the variance during training.

\section*{Acknowledgements}
We would like to express our sincere appreciation to Dr. Hang Li for his invaluable support. Concurrently, we would like to extend our gratitude to Yi Zhou for his insightful comments and guidance on this project, which provided much inspiration.

\nocite{langley00}

\bibliography{main}
\bibliographystyle{icml2024}

\onecolumn

\appendix
\setcounter{figure}{0}    
\renewcommand{\thefigure}{S\arabic{figure}}

\setcounter{table}{0}
\renewcommand{\thetable}{S\arabic{table}}

\section{Proof of Propositions \ref{prop1} and \ref{prop2}} \label{derivation}
\begin{proof}
Given $p_0(\mathbf{x}_0) = q_0(\mathbf{x}_0)e^{-k\mathcal{E} (\mathbf{x}_0)} / Z$ and $p_t(\mathbf{x}_t|\mathbf{x}_0) = q_t(\mathbf{x}_t|\mathbf{x}_0)$, the marginal distribution $p_t(\mathbf{x}_t)$ follows
\begin{align*}
    p_t(\mathbf{x}_t) & =  \int p_t(\mathbf{x}_t|\mathbf{x}_0) p_0(\mathbf{x}_0) \mathrm{~d}\mathbf{x}_0 \\
    & =  \int q_t(\mathbf{x}_t|\mathbf{x}_0) \frac{q_0(\mathbf{x}_0)e^{-k\mathcal{E} (\mathbf{x}_0)}}{Z} \mathrm{~d}\mathbf{x}_0 \\
    & =  \int q_t(\mathbf{x}_0|\mathbf{x}_t) q_t(\mathbf{x}_t) \frac{e^{-k\mathcal{E} (\mathbf{x}_0)}}{Z} \mathrm{~d}\mathbf{x}_0 \\
    & =  q_t(\mathbf{x}_t) \mathbb{E}_{q_t(\mathbf{x}_0|\mathbf{x}_t)}\left[\frac{e^{-k\mathcal{E} (\mathbf{x}_0)}}{Z} \right].
\end{align*}
Under the assumption $p_t(\mathbf{x}_t) := q_t(\mathbf{x}_t)e^{-k\mathcal{E}_t(\mathbf{x}_t)}$, $\mathcal{E}_t(\mathbf{x}_t)$ has the formula
\begin{eqnarray*}
    \mathcal{E}_t\left(\mathbf{x}_t\right)= \begin{cases}
\mathcal{E}_0\left(\mathbf{x}_0\right), & t=0 \\
-\frac{1}{k}\log \mathbb{E}_{q_{t} \left(\mathbf{x}_0|\mathbf{x}_t\right)}\left[e^{-k \mathcal{E}\left(\mathbf{x}_0\right)}\right] + \log Z, & 0<t\le 1
\end{cases}.
\end{eqnarray*}
Considering that the constant coefficient has no influence on the computation of the force field, we thus propose a notation where $\mathcal{E}_t\left(\mathbf{x}_t\right):=-\frac{1}{k}\log \mathbb{E}_{q_{t} \left(\mathbf{x}_0|\mathbf{x}_t\right)}\left[e^{-k \mathcal{E}\left(\mathbf{x}_0\right)} \right]$, $0<t\le 1$.

\par Given energy function $\mathcal{E}_t\left(\mathbf{x}_t\right)$ at time $t$, the force equation of intermediary states can be derived by
\begin{eqnarray*}
    \nabla_{\mathbf{x}_t} \mathcal{E}_t\left(\mathbf{x}_t \right) = -\frac{\int e^{-k \mathcal{E}_0(\mathbf{x}_0)}\nabla_{\mathbf{x}_t}q_t(\mathbf{x}_0|\mathbf{x}_t)\mathrm{~d}\mathbf{x}_0}{k\int q_t(\mathbf{x}_0|\mathbf{x}_t)e^{-k \mathcal{E}_0(\mathbf{x}_0)}\mathrm{~d}\mathbf{x}_0},
\end{eqnarray*}
where the numerator adopt
\begin{align*}
\int e^{-k \mathcal{E}_0(\mathbf{x}_0)}\nabla_{\mathbf{x}_t}q_t(\mathbf{x}_0|\mathbf{x}_t)\mathrm{~d}\mathbf{x}_0 & =  \int e^{- k \mathcal{E}_0(\mathbf{x}_0)}q_t(\mathbf{x}_0|\mathbf{x}_t)\nabla_{\mathbf{x}_t}\log q_t(\mathbf{x}_0|\mathbf{x}_t)\mathrm{~d}\mathbf{x}_0 \\
& =  \int q_t(\mathbf{x}_0|\mathbf{x}_t) e^{- k \mathcal{E}_0(\mathbf{x}_0)}\nabla_{\mathbf{x}_t}\log \frac{q_t(\mathbf{x}_t|\mathbf{x}_0)q_0(\mathbf{x}_0)}{q_t(\mathbf{x}_t)}\mathrm{~d}\mathbf{x}_0 \\
& =  \int q_t(\mathbf{x}_0|\mathbf{x}_t) e^{- k \mathcal{E}_0(\mathbf{x}_0)}\left(\nabla_{\mathbf{x}_t}\log q_t(\mathbf{x}_t|\mathbf{x}_0) - \nabla_{\mathbf{x}_t}\log q_t(\mathbf{x}_t) \right) \mathrm{~d}\mathbf{x}_0 \\
& = \mathbb{E}_{q_t(\mathbf{x}_0|\mathbf{x}_t)}\left[e^{- k \mathcal{E}_0(\mathbf{x}_0)}\left(\nabla_{\mathbf{x}_t}\log q_t(\mathbf{x}_t|\mathbf{x}_0) - \nabla_{\mathbf{x}_t}\log q_t(\mathbf{x}_t) \right)\right].
\end{align*}
We define $\zeta(\mathbf{x}_0, \mathbf{x}_t):=\nabla_{\mathbf{x}_t}\log q_t(\mathbf{x}_t)-\nabla_{\mathbf{x}_t}\log q_t(\mathbf{x}_t|\mathbf{x}_0)$. Therefore
\begin{align*}
\nabla_{\mathbf{x}_t} \mathcal{E}_t\left(\mathbf{x}_t \right) & =  -\frac{\mathbb{E}_{q_t(\mathbf{x}_0|\mathbf{x}_t)}\left[-e^{- k \mathcal{E}_0(\mathbf{x}_0)}\zeta(\mathbf{x}_0, \mathbf{x}_t)\right]}{k\mathbb{E}_{q_t(\mathbf{x}_0|\mathbf{x}_t)}\left[e^{- k \mathcal{E}_0(\mathbf{x}_0)}\right]} \\
& =  \frac{\mathbb{E}_{q_0(\mathbf{x}_0)}\left[q_t(\mathbf{x}_t|\mathbf{x}_0) e^{- k \mathcal{E}_0(\mathbf{x}_0)}\zeta(\mathbf{x}_0, \mathbf{x}_t)\right]}{k\mathbb{E}_{q_0(\mathbf{x}_0)}\left[q_t(\mathbf{x}_t|\mathbf{x}_0)e^{- k \mathcal{E}_0(\mathbf{x}_0)}\right]}.
\end{align*}

As $t$ approaches $0$, it prompts $q_t(\mathbf{x}_0|\mathbf{x}_t)$ involving into $\delta(\mathbf{x}_t-\mathbf{x}_0)$. Subsequently, the intermediate force turns into
\begin{align*}
    \nabla_{\mathbf{x}_t} \mathcal{E}_t\left(\mathbf{x}_t \right) & =  -\frac{\int e^{-k \mathcal{E}_0(\mathbf{x}_0)}\nabla_{\mathbf{x}_t}q_t(\mathbf{x}_0|\mathbf{x}_t)\mathrm{~d}\mathbf{x}_0}{k \int q_t(\mathbf{x}_0|\mathbf{x}_t)e^{-k \mathcal{E}_0(\mathbf{x}_0)}\mathrm{~d}\mathbf{x}_0} \\
    & =  - \frac{\mathbb{E}_{q_t(\mathbf{x}_0|\mathbf{x}_t)}\left[ e^{-k \mathcal{E}_0(\mathbf{x}_0)}/q_t(\mathbf{x}_0|\mathbf{x}_t)\cdot  \nabla_{\mathbf{x}_t}\delta(\mathbf{x}_t-\mathbf{x}_0) \right]}{k \int \delta(\mathbf{x}_t-\mathbf{x}_0)e^{-k \mathcal{E}_0(\mathbf{x}_0)}\mathrm{~d}\mathbf{x}_0} \\
    & =  - \frac{\nabla_{\mathbf{x}_t}\left[e^{-k \mathcal{E}_0(\mathbf{x}_0)} / q_t(\mathbf{x}_0|\mathbf{x}_t)\cdot q_t(\mathbf{x}_0|\mathbf{x}_t) \right]}{k e^{-k \mathcal{E}_0(\mathbf{x}_0)}} \\
    & =  \nabla_{\mathbf{x}_0} \mathcal{E}_0\left(\mathbf{x}_0 \right).
\end{align*}
As $t$ approaches $1$, both $p_t(\mathbf{x}_t)$ and $q_t(\mathbf{x}_t)$ become Gaussian distribution, prompting the energy $e^{-k\mathcal{E}_t(\mathbf{x}_t)}$ becoming a constant and force $\nabla_{\mathbf{x}_t} \mathcal{E}_t\left(\mathbf{x}_t \right)$ becoming $0$. This observation is in alignment with the theoretical deductions, as when $t\rightarrow 1$, $q_t(\mathbf{x}_t|\mathbf{x}_0)\rightarrow q_t(\mathbf{x}_t)$ and $\zeta(\mathbf{x}_0, \mathbf{x}_t)\rightarrow 0$, thus $\nabla_{\mathbf{x}_t} \mathcal{E}_t\left(\mathbf{x}_t \right)$ becomes $0$.
\end{proof}

\section{\model Architecture}\label{appendix:sde}
\subsection{Diffusion on $\mathrm{SE(3)}$} 
Diffusion models on $\mathrm{SE(3)}^N$ demonstrates powerful generative capability and are widely applied in the field of protein backbone design \citep{diffriemannian, diffusionse3}. Conducted independently over $\mathbb{R}^{3N}$ and $\mathrm{SO(3)}^N$, the diffusion process over position variables $\{\mathbf{x}_t\}_{t\in[0,1]}=[\mathbf{T}_t, \mathbf{R}_t]_{t\in[0,1]}$ are defined as
\[
\mathrm{d} \mathbf{x}_t=\left[-\frac{1}{2} \beta_t \mathbf{P} \mathbf{T}_t, 0\right] \mathrm{d} t+\left[\sqrt{\beta_t} \mathbf{P} \mathrm{d} \mathbf{w}_t, \sqrt{\frac{\mathrm{d}}{\mathrm{d} t} \sigma^2_t} \mathrm{d} \mathbf{w}^{\mathrm{SO}(3)}_t\right], \quad \mathbf{x}_0\sim q_0(\mathbf{x}_0)
\]
where $[\mathbf{w}_t, \mathbf{w}_t^{\mathrm{SO(3)}}]$ is the standard Wiener process in $[\mathcal{N}(0, I_3)^{\otimes N} , \mathcal{U}(\mathrm{SO}(3))^{\otimes N}]$. Here, $\beta_t=\beta_{\min}+t(\beta_{\max}-\beta_{\min})$ and we set $\beta_{\min}=0.1$, $\beta_{\max}=20$. Similarly, $\sigma(t)=\log(te^{\sigma_{\max}}+(1-t)e^{\sigma_{\min}})$ and we set $\sigma_{\min}=0.1$, $\sigma_{\max}=1.5$. Adding noise according to the aforementioned equation, the transition kernel satisfies $q_t(\mathbf{T}_t|\mathbf{T}_0)=\mathcal{N}(\mathbf{T}_t; \sqrt{\alpha_t} \mathbf{T}_0,(1-\alpha_t)I)$, where $\alpha_t:=\exp{\left(-\int_0^t \beta(s)\mathrm{~d}s\right)}$, and the rotation kernel satisfies $p_t(\mathbf{R}_t|\mathbf{R}_0)=\mathrm{IGSO}_3(\mathbf{R}_t;\mathbf{R}_0, \sigma^2_t)$,
\begin{eqnarray*}
\mathrm{IGSO}_3(\mathbf{R}_t; \mathbf{R}_0, \sigma^2_t) = \sum_{\ell}(2 \ell+1) \mathrm{e}^{-\frac{\ell(\ell+1) \sigma^2_t}{2}} \frac{\sin ((\ell+\frac{1}{2}) \omega_t)}{\sin (\frac{\omega_t}{2})},
\end{eqnarray*}
where $\omega_t=\mathrm{Axis\_angle}(\mathbf{R}_0^{\top}\mathbf{R}_t)$. Correspondingly, the associated reverse time diffusion process follows
\[
\mathrm{d} \mathbf{x}_t=\left[\mathbf{P}\left(-\frac{1}{2} \beta_t  \mathbf{T}_t-\beta_t\nabla\log q_t(\mathbf{T}_t)\right), -\frac{\mathrm{d}}{\mathrm{d} t} \sigma^2_t \nabla\log q_t(\mathbf{R}_t) \right] \mathrm{d} t+\left[\sqrt{\beta_t} \mathbf{P} \mathrm{d} \bar{\mathbf{w}}_t, \sqrt{\frac{\mathrm{d}}{\mathrm{d} t} \sigma^2_t} \mathrm{d} \bar{\mathbf{w}}^{\mathrm{SO}(3)}_t\right].
\]
Considering the intractable marginal score function $\nabla\log q_t(\mathbf{x}_t)=[\nabla\log q_t(\mathbf{T}_t) ,\nabla\log q_t(\mathbf{R}_t)]$, we use score matching techniques to estimate them respectively. Training the neural networks $s_{\theta}(\mathbf{T}_t, t)$ and $s_{\theta}(\mathbf{R}_t, t)$, the denoising score matching loss function is defined as
\begin{align*}
    \mathcal{L}_{\mathrm{DSM}}^{\mathbf{T}} & = \mathbb{E}_{p(t)}\mathbb{E}_{q_0(\mathbf{T}_0), q_t(\mathbf{T}_t|\mathbf{T}_0)}\left[\lambda(t) \| s_{\theta}(\mathbf{T}_t, t) - \nabla_{\mathbf{T}_t}\log p_t(\mathbf{T}_t|\mathbf{T}_0)\|^2 \right], \\
    \mathcal{L}_{\mathrm{DSM}}^{\mathbf{R}} & =  \mathbb{E}_{p(t)}\mathbb{E}_{q_0(\mathbf{R}_0), q_t(\mathbf{R}_t|\mathbf{R}_0)}\left[\lambda^r(t) \| s_{\theta}^r(\mathbf{R}_t, t) - \nabla_{\mathbf{R}_t}\log p_t(\mathbf{R}_t|\mathbf{R}_0)\|^2 \right],
\end{align*}
where $\lambda(t)$, $\lambda^r(t)$ is a time weighting function and the conditional score has the following formula 
\begin{align*}
    \nabla_{\mathbf{T}_t}\log p_t(\mathbf{T}_t|\mathbf{T}_0) & =  -\frac{\mathbf{T}_t - \sqrt{\alpha_t}\mathbf{T}_0}{1-\alpha_t}, \\
    \nabla_{\mathbf{R}_t}\log p_t(\mathbf{R}_t|\mathbf{R}_0) & =  \frac{\mathbf{R}_t}{\omega_t} \log \left(\mathbf{R}_0^{\top}\mathbf{R}_t\right) {\partial_{\omega_t} \log \mathrm{IGSO}_3\left(\mathbf{R}_t; \mathbf{R}_0, \sigma^2_t\right)}.
\end{align*}
To improve the quality of local atomic structures, auxiliary losses are added to the optimization problem, including the MSE loss on the backbone (bb) positions $\mathcal{L}_{\mathrm{bb}}$ and a local neighborhood loss on pairwise atomic distances $\mathcal{L}_{\mathrm{2D}}$. The total loss can be written as
\[
\mathcal{L} = \mathcal{L}_{\mathrm{DSM}}^{\mathbf{T}} + 0.5*\mathcal{L}_{\mathrm{DSM}}^{\mathbf{R}} + 0.25* \Huge{1} \left\{t<\frac{\mathrm{T}_{\mathrm{F}}=0.25}{4}\right\}\left(\mathcal{L}_{\mathrm{bb}}+\mathcal{L}_{\mathrm{2D}} \right).
\]
\subsection{Interpolation with Simulation Data}\label{appendix:interpolation} 
According to the derivation in Appendix~\ref{derivation}, $\nabla_{\mathrm{x}_0}\mathcal{E}_0(\mathrm{x}_0)$ has the ground truth value at the boundary $t\in\{ 0, 1\}$ and we observe that both score functions 
$\nabla_{\mathbf{x}_t}\log q_t(\mathbf{x}_t|\mathbf{x}_0)$ and $\nabla_{\mathbf{x}_t}\log q_t(\mathbf{x}_t)$
exhibit large variance at $t=0$ (due to small noise level at small $t$). 
Thus, we propose an interpolation strategy for the intermediate force function like in \citep{mate2023learning, wang2024energy}:
\begin{eqnarray*}
    h_{\psi}(\mathbf{x}_t, t) = (1-t)\nabla_{\mathbf{x}_0} \mathcal{E}_0(\mathbf{x}_0) + t(1-t) g_{\psi}(\mathbf{x}_t, t).
\end{eqnarray*}
where $g_{\psi}(\mathbf{x}_t, t)$ is a neural network to estimate the intermediate term of interpolation construction. It ensures that as $t$ approaches $0$, the network approximates the ground truth force $\nabla_{\mathbf{x}_0} \mathcal{E}_0(\mathbf{x}_0)$ and as $t$ nears $1$, the network becomes $0$.
\[
h_{\psi}(\mathbf{x}_t, t) = \begin{cases}
\nabla_{\mathbf{x}_0} \mathcal{E}_0(\mathbf{x}_0), & t=0 \\
\mathrm{Parametrized\ Interpolation}, & 0<t<1 \\
0, & t=1
\end{cases}.
\]


Nevertheless, throughout the inference process, despite the ability to derive a predicted $\hat{\mathbf{x}}_0$ at every step for the computation of $\nabla_{\hat{\mathbf{x}}_0}\mathcal{E}_0(\hat{\mathbf{x}}_0)$, the utilization of OpenMM for energy calculation is highly time-consuming, particularly in estimating energy in lengthy protein chains. To further demonstrate the computational cost of potential energy evaluation, we tabulate the total time consumption of once energy/force calculation using openMM (the cost of side-chain packing using faspr is excluded) for the fast-folding proteins. Due to the 1000 steps in the inference diffusion process, the time required to generate a single protein conformation is multiplied by 1000.
\begin{table}[h]
\centering
\caption{The time consumption of once energy/force calculation using openMM.}
\vskip 0.1in
\begin{sc}
\begin{tabular}{ccccccc}
\toprule
 & 1FME & 2F4K & 2JOF & 2WAV & A3D & CLN025 \\ 
\midrule
{Length} & 28 & 35 & 20 & 47 & 73 & 10 \\ 
{CPU} & 3.6±0.6 s & 4.2±1.8 s & 1.4±0.3 s & 8.3±3.3 s & 19.4±5.3 s & 0.9±0.2 s \\ 
{GPU} & 1.7±0.2 s & 1.7±0.5 s & 1.6±0.5 s & 1.9±0.6 s & 2.2±0.5 s & 1.6±0.5 s \\ 
\bottomrule
\toprule
\multicolumn{1}{l}{{}} & GTT & NTL9 & NuG2 & PRB & UVF & lambda \\
\midrule
{Length} & 35 & 39 & 56 & 47 & 52 & 80 \\ 
{CPU} & 5.0±2.1 s & 6.8±1.4 s & 12.1±5.5 s & 8.7±1.6 s & 12.6±2.5 s & 26.6±16.4 s \\
{GPU} & 1.8±0.5 s & 1.8±0.6 s & 2.0±0.5 s & 1.8±0.5 s & 2.0±0.5 s & 2.0±0.5 s \\ 
\bottomrule
\end{tabular}
\end{sc}
\end{table}
\vskip 0.1in
Considering that the reverse diffusion process is discretized into 1000 steps, the time expenditure is not feasible to incorporate the ground-truth MD force field in each reverse sampling step, necessitating the employment of a surrogate model. Thus, we additionally train a network $\tilde{g}_{\nu}(\mathbf{x}_t, t)$ to estimate the ground truth force field $\nabla_{\mathbf{x}_0} \mathcal{E}_0(\mathbf{x}_0)$ at the initial moment by MSE loss function $\mathcal{L}_{t=0}(\nu)$, resulting in the transformation of $h_{\psi}(\mathbf{x}_t, t)$ into
\begin{eqnarray}\label{eq:interpolation_net}
    h_{{\psi},\nu}(\mathbf{x}_t, t)=(1-t)\tilde{g}_{\nu}({\mathbf{x}}_t, t) + t(1-t) g_{\psi}(\mathbf{x}_t, t).
\end{eqnarray}
The optimization problem, combining two loss functions, is defined as:
\begin{eqnarray*}
    \min_{\psi,\nu}\mathcal{L}(\psi,\nu) = \min_{\psi, \nu}\left[ \mathcal{L}_{\mathrm{force}}(\psi,\nu) + \mathcal{L}_{t=0}(\nu)\right],
\end{eqnarray*}
where
\begin{eqnarray*}
  &&\mathcal{L}_{\mathrm{force}}(\psi,\nu) = \mathbb{E}_{p(t),\mathbf{x}_0, \mathbf{x}_t}\bigg[ \bigg\| h_{{\psi},\nu}(\mathbf{x}_t, t) - \frac{e^{-k \mathcal{E}_0(\mathbf{x}_0)}\zeta(\mathbf{x}_0, \mathbf{x}_t)}{\mathbb{E}_{\mathbf{x}_0^K}\left[q_t(\mathbf{x}_t|\mathbf{x}_0) e^{-k \mathcal{E}_0(\mathbf{x}_0)}\right]} \bigg\|^2_2 \bigg] \\
  &&\mathcal{L}_{t=0}(\nu) = \mathbb{E}_{p(t)}\mathbb{E}_{\mathbf{x}_0, \mathbf{x}_t}\left[\|\tilde{g}_{\nu}(\mathbf{x}_t, t)- \nabla_{\mathbf{x}_0}\mathcal{E}_0(\mathbf{x}_0)\|^2_2\right].
\end{eqnarray*}

\vspace{3mm}
\subsection{Implementation}
\label{implement}
\begin{figure*}[h]
    \centering
    \includegraphics[width=1.0\linewidth]{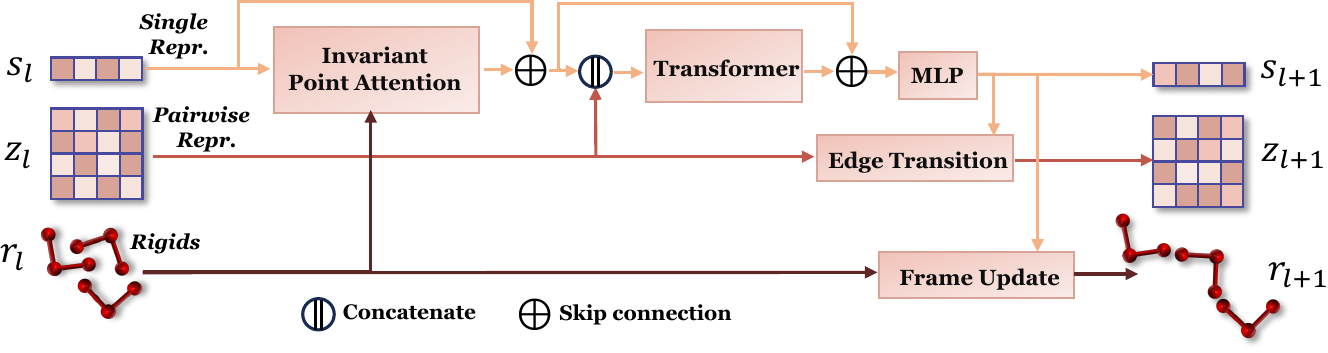}
    \caption{single-layer of \model. This architecture, drawing references from \citet{alphafold,diffusionse3}, can be widely applied in the unconditional model, the conditional model, the intermediate energy prediction network $f_{\phi}(\mathbf{x}_t, t)$, the intermediate force prediction network ${g}_{\nu}(\mathbf{x}_t, t)$ and the force-field prediction network $\tilde{g}_{\nu}(\mathbf{x}_t, t)$ used in Eq.~\ref{eq:interpolation_net}. The invariant point attention, edge transition, and frame update modules similar to the corresponding structures in \citet{alphafold}}
    \label{fig:ipa}
\end{figure*}

The single layer of \model consists of invariant point attention layer~\citep{alphafold}, transformer and other structure modules similar to ~\citet{diffusionse3}.
In the unconditional diffusion model, the single representation $s_{0}$ is initialized as a concatenation of the position encoding of residues and sinusoidal time embedding. Meanwhile, the pair representation $z_{0}$ is initialized as a concatenation of the relative residue position encoding, sinusoidal time embedding, and relative sequence distances. In the conditional model, the node embedding of ESMfold~\citet{esmfold} with corresponding sequences is additionally concatenated to the single representation $s_{0}$. In addition, we apply the same noise-injection schedule as the translation term in Eq.~(\ref{eq:sde-r3-forward}) to the ESMFold node embeddings before feeding to the neural networks. This noise injection step further increases sampling diversity. In both the intermediate force prediction network ${g}_{\nu}(\mathbf{x}_t, t)$ and the force-field prediction network $\tilde{g}_{\nu}(\mathbf{x}_t, t)$, the features are initialized as the output single and pair representation of the conditional model. These features are then used to predict the intermediate force, which is derived from the last three dimensions of the newly predicted rigids. The intermediate energy prediction network 
$f_{\phi}(\mathbf{x}_t, t)$ also takes the output of the conditional model as its input and a linear layer is added at the end that transforms the single representation into a scalar value, thereby aiming to predict energy.
Detailed hyperparameters are listed in \ref{tab:hparam}.

\subsection{Computational Cost}
Our \model-base model has similar computational complexity to other diffusion-based conformation generation models, e.g., FrameDiff \citep{diffusionse3} and Str2Str \citep{str2str}. The introduction of a guidance network will not significantly increase computational cost. Specifically, the energy/force guidance modules only introduce a small computational overhead during inference through a \textbf{lightweight energy/force prediction head}, which is based on the pretrained \model-base model to accomplish guidance.

We made a comparative study of the time and GPU memory requirement for generating protein conformations across different sequence lengths. We used a single NVIDIA-V100 GPU to benchmark model performance. The batch size is 16, and the number of reverse sampling steps is 1000.

\begin{table}[h]
\centering
\caption{The time and memory requirement for generating protein conformations across different sequence lengths.}
\vskip 0.1in
\begin{sc}
\begin{tabular}{lcccccc}
\toprule
\multicolumn{1}{c}{} & \multicolumn{2}{c}{\model-base} & \multicolumn{2}{c}{\model-Energy} & \multicolumn{2}{c}{\model-Force} \\ 
\midrule
len (protein) & \multicolumn{1}{c}{Time (sec)} & Memory (MB) & \multicolumn{1}{c}{Time (sec)} & Memory (MB) & \multicolumn{1}{c}{Time (sec)} & Memory (MB) \\ 
\midrule
58 (BPT) & \multicolumn{1}{c}{71} & 1838 & \multicolumn{1}{c}{110} & 3142 & \multicolumn{1}{c}{104} & 1916 \\
118 (1JM4.B) & \multicolumn{1}{c}{167} & 3932 & \multicolumn{1}{c}{256} & 9722 & \multicolumn{1}{c}{206} & 4010 \\
238 (2LAO.A) & \multicolumn{1}{c}{578} & 12402 & \multicolumn{1}{c}{888} & 29886 & \multicolumn{1}{c}{709} & 12480 \\ 
\bottomrule
\end{tabular}
\end{sc}
\end{table}
\vskip 0.1in
Notably, the energy guidance policy almost doubles the memory cost compared to \model-base, since the derivative of energy is required to compute the force guidance term. Meanwhile, force guidance only consumes negligibly more computational resources compared to \model-base, which is another significant advantage of the force guidance strategy versus the energy guidance strategy.

\vspace{3mm}
\newpage
\subsection{Hyperparameters}


 \begin{table*}[hbt!]
 \label{tab:hparam}
\centering
\caption{Hyperparameter choices of \model}
\setlength{\tabcolsep}{16mm
}
\begin{tabular}{ll}
\toprule[1.2pt]

\textbf{Hyperparameters} & \textbf{Values}\\ 
 \midrule[0.5pt]
 \multicolumn{2}{c}{\textbf{Neural network}} \\
 \midrule[0.5pt]
Number of IPA blocks & 4 \\
Dimension of single repr. & 256 \\
Dimension of pairwise Repr.& 128 \\
Dimension of hidden & 256 \\
Number of IPA attention heads & 4 \\
Number of IPA query points &  8 \\
Number of IPA value points &  12 \\
Number of transformer attention heads&4 \\
Number of transformer layers&  2 \\

  \midrule[0.5pt]
 \multicolumn{2}{c}{\textbf{SDE}} \\
 \midrule[0.5pt]
  Number of time steps &  1000 \\
  Translation scheduler &  Linear \\
  Translation $\beta_{\min}$ &  0.1 \\
  Translation $\beta_{\max}$ &  20 \\
  Rotation scheduler &  Logarithmic \\
  Rotation $\sigma_{\min}$ &  0.1 \\
  Rotation $\sigma_{\max}$ &  1.5 \\
 
\midrule[0.5pt]
\multicolumn{2}{c}{\textbf{Training}} \\
\midrule[0.5pt]
Batch Size  &  32 \\
Learning Rate   & 1$\times 10^{-4}$  \\
Optimizer  &  Adam (weight decay = $0.01$) \\
Learning Rate Scheduler & ReduceLROnPlateau (factor = $0.5$ )  \\
Warm up steps  &  5000 \\
\midrule[0.5pt]
\multicolumn{2}{c}{\textbf{Inference}} \\
\midrule[0.5pt]
 Classifier-free guidance strength $\gamma$ &  $\left[0.5,0.8,1.0\right]$ \\
 Force guidance strength $\eta$ &  $\left[0.5,0.7,1.0,1.3\right]$ \\

\bottomrule[1.2pt]
\end{tabular}

\label{tab:hyperparameters}
\end{table*}

\clearpage

\section{Experimental Details}

\subsection{Datasets}
\label{si:datasets}

\textbf{Training datasets.} We train \model using all available single-chain protein structures deposited to the PDB on or before Dec 31, 2021, and perform validation using structures deposited between Jan 1, 2022 and Dec 31, 2022. Unlike previous processing pipelines which typically only extract the first model (i.e., only take ``MODEL 1") of each chain, we parse and save all available models in order to make full use of available experimental structures, especially the solution NMR data. We train and validate on protein chains with residue length between 20 and 400, and remove data with less than 70\% valid backbone atoms, which leads to 412,371 (57,088) training (validation) structures. During training, we adopt a 30\% sequence-similarity-based clustering strategy within our data sampler, where each data loader first randomly samples a cluster, then subsamples a structure from that cluster.

\textbf{Fast-folding proteins.} The simulation details of fast-folding proteins are described in \cite{deshawfastfold2011} and the trajectories are obtained from the authors. The dataset contains 12 small proteins range from 10 to 80 residues that exhibit multiple fast folding-unfolding events within the millisecond-scale all-atom simulations. Atom coordinates are saved at 200~\si{\pico\second} intervals and 520K $\sim$ 14.7M aggregated frames are collected for each protein. $\alpha$-carbon atom coordinates are used for analysis. For sampling, we generate 1000 conformations for each proteins with guidance strength $\eta = 1.3$ for both \textit{force} and \textit{energy guidance}, and the sequence condition $\gamma$ sampled between $0.5 \sim 1.0$.

\textbf{BPTI.} Bovine pancreatic trypsin inhibitor (BPTI) is a 58-residue model protein for protein dynamics. Transitions between five meta-stable states are identified in a previous 1 \si{\milli\second} simulation study \cite{deshawbpti2010}. MD trajectories are obtained from the authors and contain atom coordinates saved at 250 \si{\pico\second} intervals with 4.1M total frames. $\alpha$-carbon atom coordinates are used for analysis. The coordinate PDB files for the five kinetic clusters are provided in the supplementary materials for \cite{deshawbpti2010}. For sampling, we generate 1000 conformations for BPTI with guidance strength $\eta = 1.5$ for \textit{force guidance} and $0.5$ for \textit{energy guidance}. The sequence condition $\gamma$ is set to sample between $0.8\sim 1.0$.

\subsection{Time-Lagged Independent Analysis (TICA)}
\label{mtd:tica}

For fast-folding proteins and BPTI, we use \texttt{pyemma} \cite{pyemma_2015} for featurizing protein structures (as pairwise distances between $\alpha$-carbons), and use \texttt{Deeptime} \cite{hoffmann2021deeptime} for time-lagged independent analysis (TICA) and projecting the protein conformations into the two slowest components for distribution comparison \cite{naritomi2013tica, perez2013tica}.

\subsection{Energy and Force Evaluation}
\label{mtd:openmm}

Potential energy of protein structures are evaluated using \texttt{OpenMM} \cite{eastman2017openmm} following a similar pipeline as in \cite{str2str}. First, we use \texttt{faspr} \cite{huang2020faspr} for side-chain packing (i.e., adding side chain atoms with optimized coordinates) on generated backbone conformations. Protonation (i.e., adding hydrogen atoms), solvation, and energy minimization are performed using \texttt{OpenMM} with Amber-ff14SB force field \cite{maier2015amber} and implicit solvent GBn2 \cite{nguyen2013gbn2}. During minimization, we keep the conformations by applying independent harmonic restraints on all heavy (non-hydrogen) atoms with spring constant of 10 \si{\kilo cal\per \mol \cdot \angstrom^2}, the tolerance is set to 2.39 \si{\kilo cal\per \mol \cdot \angstrom^2} without maximal step limits. The potential energy excluding the component from harmonic restraints is reported (unit: \si{\kilo cal\per \mol}). Forces on each $\alpha$-carbon atoms are also reported for training the neural networks used in \textit{force guidance}.

\subsection{Details on Evaluation Metrics}
\label{si:metrics}

\par \textbf{Jensen-Shannon (JS) distance between sample distributions.} We measure the similarity between two conformation ensembles by JS distance:

\begin{eqnarray*}
    \mathrm{JS}(p\|q) = \sqrt{\frac{\mathrm{KL}(p\|q) + \mathrm{KL}(q\|p)}{2}},
\end{eqnarray*}
where $\mathrm{KL}(\cdot\|\cdot)$ is the Kullback-Leibler divergence. Similar to \cite{str2str}, we project coordinates to three types of feature spaces to compute conformation distribution: 1) Pairwise Distance (PwD) between $\alpha$-carbon of residues excluding adjacent residues within an offset of 3; 2) Radius of gyrations (Rg) that are distances of $\alpha$-carbon atoms to the center-of-mass of the conformation; 3) Time-lagged Independent Components (TIC) from TICA analysis on the reference MD trajectories (see Appendix \ref{mtd:tica}), only the first two components are used for JS calculation. Feature dimensions are discretized with 50 bins according to the values of the reference MD samples and a pseudo count of $10^{-6}$ is added. We compute the distance over each dimension and report the mean values as the final JS distance.

The first two dimensions of TIC represent the a low dimensional 2D space with features most predictive for protein dynamics, therefore we also estimate the sample distribution of the joint space of these two dimensions, labeled as TIC-2D.

\par \textbf{Contact rates between residues.} For fast-folding proteins we compute the contact rate between a pair of residues as the fraction of conformation with contacting residues for each pair of residues at a threshold of 10 \si{\angstrom}:
\begin{eqnarray*}
  r(u, v) = \frac{1}{N} \sum_{i = 1}^{N} \mathbf{1} \{d_i(u, v) \le 10 \si{\angstrom}\},
\end{eqnarray*}
where $d_i(u, v)$ is the Euclidean distance between the $\alpha$-carbon of residue $u$ and $v$, $N$ is the number of samples, and $\mathbf{1}\{\cdot \}$ is the indicator function.

To quantify the accuracy of predicting the residue contacts, we compute the root-mean-square error between generated samples and reference MD samples:
\begin{eqnarray*}
    \text{RMSE}_\text{contact} = \sqrt{\frac{2}{L(L-1)}\sum_{u=1}^{L-1}\sum_{v=u+1}^{L} \big( r(u, v) - r_\text{ref}(u, v)\big)^2 }.
\end{eqnarray*}

\par \textbf{Validity.}
We adopt the \textit{Validity} metrics in \cite{str2str} to evaluate the backbone structural validity of generated conformations. A conformation is consider to have \textit{Clash} if the distance between any two $\alpha$-carbon is less than two times the van der Waals radius of $\alpha$-carbon with an overlap tolerance: $\delta  = 2\times 1.7 -  0.4 = 3.0 $ \si{\angstrom}. For \textit{bond break}, we surveyed the maximum adjacent $\alpha$-carbon distances in the reference MD trajectories of fast-folding proteins and BPTI. We found the maximum distance is around 4.19 \si{\angstrom} consistent cross different protein and therefore use it as a cutoff value. Any conformation contains distance between adjacent $\alpha$-carbon greater than 4.19 \si{\angstrom} is consider having \textit{bond break} and an invalid structure. For a rigorous evaluation, we report the overall validity \textsc{Val-CA} as the fraction of conformations without \textit{Clash} and \textit{Bond break}.

\clearpage

\section{Additional Experimental Results}

\subsection{Ablation Studies on Force and Energy Guidance}

\begin{figure*}[h]
    \vskip 0.2in
    \centering
    \includegraphics[width=0.7\linewidth]{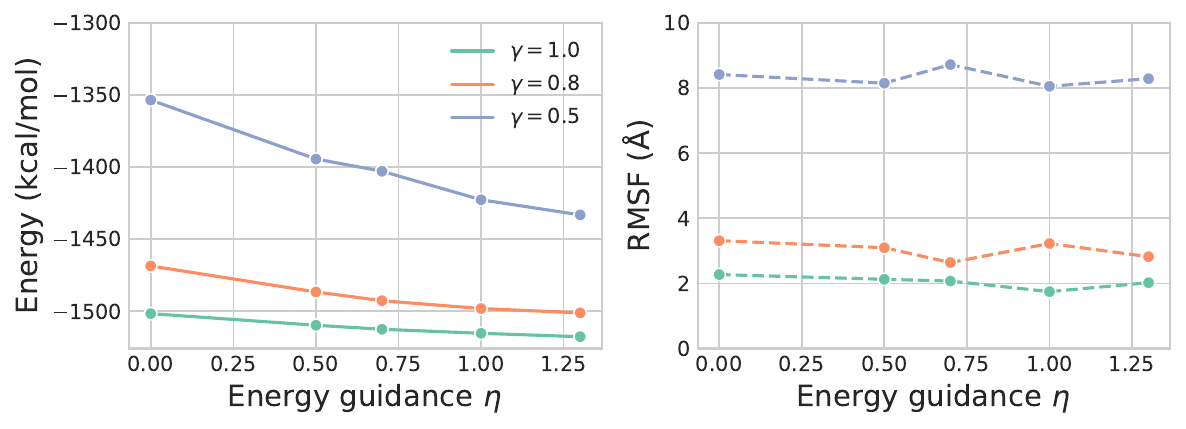}
    
    \caption{Energy (left) and diversity (RMSF, right) of sampled conformation for WW domain with various levels of energy guidance ($\eta$) and sequence condition ($\gamma$).}
    \label{fig:energy-guide-eta}
\end{figure*}

In addition to different levels of \textit{force guidance} shown in the main text, conducted similar ablation study on \textit{energy guidance} and on other proteins in fast-folding protein datasets. As shown in Figure~\ref{fig:energy-guide-eta}, similar to \textit{force guidance}, \textit{energy guidance} also improve sample energy while maintaining the level of sample diversity. 

In addition, we compare the fast-folding related metrics in Table \ref{tab:ablation}. Notably, validity score (\textsc{Val-CA}) improves after adding energy and force guidance.

\begin{table*}[htp]
\caption{Ablation studies on different levels of force and energy guidance on \model, performance is evaluated on fast-folding proteins. mean/median values are reported.}
\label{tab:ablation}
\vskip 0.15in
\begin{center}
\begin{small}
\begin{sc}
\begin{tabular}{lcccccccc}
\toprule
                        & \multicolumn{4}{c}{J-S distance ($\downarrow$)} &  Val-CA         & RMSE$_\text{contact}$ & RMSF  \\ \cmidrule(lr){2-5}
Models                  & PwD   &  Rg               & TIC    & TIC-2D     & ($\uparrow$)    & ($\downarrow$)  &  (\si{\angstrom})    \\
\midrule
\multicolumn{9}{c}{\model-Energy} \\
\midrule
$\gamma = 1.0, \eta=0.0$ & 0.41/0.44 & 0.40/0.44 & 0.47/0.51 & 0.59/0.61 & 0.97/0.98 & 4.87/4.69 & 2.3/2.0 & \\
$\gamma = 1.0, \eta=0.5$ & 0.39/0.39 & 0.37/0.38 & 0.45/0.51 & 0.58/0.62 & 0.97/0.99 & 4.75/4.27 & 3.3/2.6 & \\
$\gamma = 1.0, \eta=0.7$ & 0.39/0.41 & 0.37/0.40 & 0.45/0.51 & 0.58/0.62 & 0.97/0.99 & 4.69/4.32 & 3.7/3.0 & \\
$\gamma = 1.0, \eta=1.0$ & 0.39/0.41 & 0.37/0.38 & 0.45/0.50 & 0.58/0.62 & 0.98/0.99 & 4.62/4.33 & 4.0/3.0 & \\
$\gamma = 1.0, \eta=1.3$ & 0.40/0.39 & 0.37/0.36 & 0.45/0.49 & 0.58/0.61 & 0.98/0.99 & 4.56/4.33 & 4.5/3.4 & \\
\midrule
$\gamma = 0.8, \eta=0.0$ & 0.33/0.33 & 0.31/0.31 & 0.45/0.46 & 0.61/0.63 & 0.92/0.93 & 4.09/4.10 & 4.2/4.0 & \\
$\gamma = 0.8, \eta=0.5$ & 0.34/0.36 & 0.31/0.32 & 0.41/0.44 & 0.56/0.60 & 0.96/0.98 & 4.10/3.98 & 4.8/4.6 & \\
$\gamma = 0.8, \eta=0.7$ & 0.35/0.36 & 0.32/0.32 & 0.42/0.44 & 0.56/0.60 & 0.97/0.98 & 4.09/4.05 & 5.0/4.4 & \\
$\gamma = 0.8, \eta=1.0$ & 0.36/0.35 & 0.32/0.31 & 0.42/0.44 & 0.56/0.60 & 0.97/0.98 & 4.07/4.09 & 5.3/5.1 & \\
$\gamma = 0.8, \eta=1.3$ & 0.36/0.37 & 0.33/0.34 & 0.42/0.43 & 0.56/0.60 & 0.97/0.98 & 4.04/4.08 & 5.8/5.2 & \\
\midrule
$\gamma = 0.5, \eta=0.0$ & 0.29/0.26 & 0.26/0.23 & 0.41/0.41 & 0.57/0.62 & 0.84/0.88 & 3.48/3.68 & 8.2/8.3 & \\
$\gamma = 0.5, \eta=0.5$ & 0.33/0.33 & 0.31/0.29 & 0.41/0.40 & 0.57/0.60 & 0.92/0.96 & 3.51/3.16 & 8.7/8.3 & \\
$\gamma = 0.5, \eta=0.7$ & 0.35/0.33 & 0.33/0.30 & 0.41/0.40 & 0.57/0.60 & 0.94/0.97 & 3.60/3.26 & 8.6/8.7 & \\
$\gamma = 0.5, \eta=1.0$ & 0.36/0.34 & 0.35/0.31 & 0.41/0.39 & 0.57/0.60 & 0.95/0.96 & 3.68/3.42 & 9.0/9.2 & \\
$\gamma = 0.5, \eta=1.3$ & 0.37/0.35 & 0.36/0.33 & 0.41/0.40 & 0.58/0.60 & 0.95/0.96 & 3.76/3.51 & 8.9/9.1 & \\
\midrule
\multicolumn{9}{c}{\model-Force} \\
\midrule
$\gamma = 1.0, \eta=0.0$ & 0.41/0.44 & 0.40/0.44 & 0.47/0.51 & 0.59/0.61 & 0.97/0.98 & 4.87/4.69 & 2.3/2.0 \\
$\gamma = 1.0, \eta=0.5$ & 0.38/0.42 & 0.36/0.41 & 0.46/0.50 & 0.58/0.64 & 0.97/0.98 & 4.66/4.27 & 3.3/3.1 \\
$\gamma = 1.0, \eta=0.7$ & 0.38/0.42 & 0.36/0.40 & 0.46/0.50 & 0.58/0.63 & 0.97/0.98 & 4.58/4.22 & 3.3/3.0 \\
$\gamma = 1.0, \eta=1.0$ & 0.38/0.42 & 0.36/0.40 & 0.46/0.51 & 0.59/0.64 & 0.98/0.99 & 4.46/4.11 & 3.3/2.8 \\
$\gamma = 1.0, \eta=1.3$ & 0.38/0.41 & 0.36/0.40 & 0.47/0.52 & 0.59/0.64 & 0.98/0.99 & 4.37/3.97 & 3.3/2.8 \\
\midrule
$\gamma = 0.8, \eta=0.0$ & 0.33/0.33 & 0.31/0.31 & 0.45/0.46 & 0.61/0.63 & 0.92/0.93 & 4.09/4.10 & 4.2/4.0 \\
$\gamma = 0.8, \eta=0.5$ & 0.32/0.30 & 0.30/0.27 & 0.42/0.46 & 0.56/0.59 & 0.95/0.96 & 3.87/3.64 & 4.6/4.3 \\
$\gamma = 0.8, \eta=0.7$ & 0.32/0.29 & 0.30/0.26 & 0.42/0.45 & 0.57/0.60 & 0.96/0.97 & 3.82/3.60 & 4.5/4.2 \\
$\gamma = 0.8, \eta=1.0$ & 0.32/0.29 & 0.29/0.26 & 0.43/0.46 & 0.58/0.61 & 0.97/0.98 & 3.71/3.51 & 4.6/4.4 \\
$\gamma = 0.8, \eta=1.3$ & 0.32/0.30 & 0.30/0.28 & 0.43/0.46 & 0.58/0.60 & 0.98/0.98 & 3.67/3.58 & 4.7/4.1 \\
\midrule
$\gamma = 0.5, \eta=0.0$ & 0.29/0.26 & 0.26/0.23 & 0.41/0.41 & 0.57/0.62 & 0.84/0.88 & 3.48/3.68 & 8.2/8.3 \\
$\gamma = 0.5, \eta=0.5$ & 0.31/0.28 & 0.28/0.25 & 0.41/0.41 & 0.58/0.62 & 0.91/0.95 & 3.38/3.32 & 8.5/8.6 \\
$\gamma = 0.5, \eta=0.7$ & 0.31/0.28 & 0.28/0.26 & 0.41/0.41 & 0.58/0.62 & 0.93/0.97 & 3.32/3.24 & 8.3/8.7 \\
$\gamma = 0.5, \eta=1.0$ & 0.31/0.29 & 0.29/0.26 & 0.42/0.41 & 0.58/0.63 & 0.96/0.97 & 3.45/3.27 & 8.5/9.1 \\
$\gamma = 0.5, \eta=1.3$ & 0.32/0.30 & 0.30/0.28 & 0.41/0.40 & 0.58/0.63 & 0.97/0.98 & 3.52/3.29 & 8.4/8.5 \\
\midrule
\bottomrule
\end{tabular}
\end{sc}
\end{small}
\end{center}
\vskip 3mm
\end{table*}

\newpage
\subsection{Additional Discussion on Fast-folding Results}

\begin{figure*}[h]
    \vskip 0.2in
    \centering
    \includegraphics[width=0.8\linewidth]{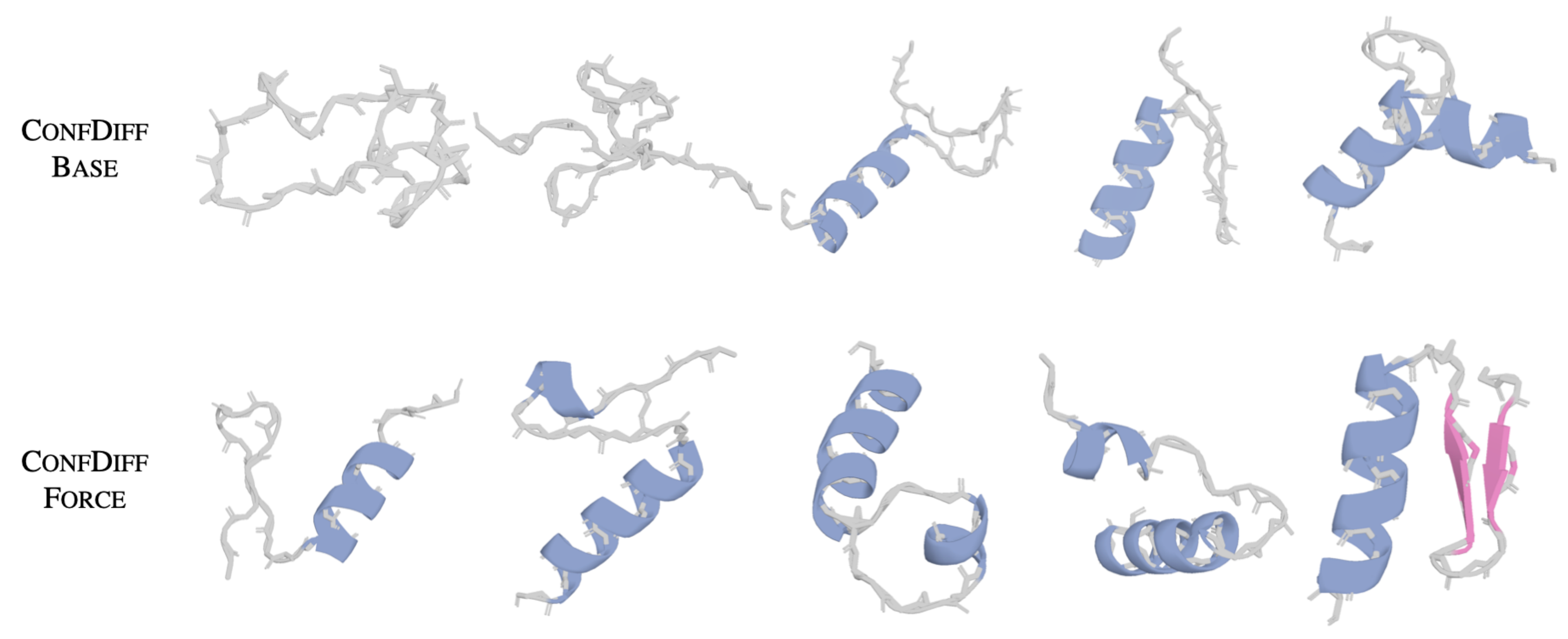}
    \caption{Five worst structures with highest residue clash rate. Both the cartoon representation and backbone bonds are shown to better depict the clashes.}
    \label{fig:worst-structures}
\end{figure*}

We find that Jensen-Shannon (JS) metrics alone are insufficient to reflect the quality and diversity of sampled conformations. Note that the JS distance here is calculated and averaged over each dimension (e.g., $\alpha$-carbons), which is not able to capture the bond breaking and steric clash. With the introduction of energy/force guidance, irrational conformations are rectified and structures tend to contract towards the local minimum. This is the main factor leading to a performance drop in distributional metrics.

Empirically, \model-base sometimes samples poor structures with steric clashes, while energy/force guidance strategies enable sampling high-fidelity conformations (i.e., high VAL-CA scores) with slightly worse JS metrics. As shown in Figure \ref{fig:energy-vs-eta} in our paper, RMSF (reflecting structural diversity) values fluctuate with various force guidance strengths, while the potential energy keeps improving. This also explains the fluctuation of JS metrics with improved VAL-CA scores after energy/force guidance.

Here we show some examples to compare sampled structures from \model-base and \model-force. Among 1000 samples from \model-base ($\gamma=0.5$) and \model-force ($\gamma=0.5$, $\eta=1.3$), we select five worst conformations (with highest clash rate) shown in the Figure \ref{fig:worst-structures}. \model-force samples exhibit fewer steric clashes, demonstrating more realistic protein structures, despite slightly worse JS distance. Therefore, it is more appropriate to take into account all metrics for a comprehensive assessment of the generated conformation ensemble.

\clearpage

\subsection{Sample Distribution over TIC Components for All Fast-Folding Proteins}

Here we include sample distributions over the first two TIC components for all twelve fast-folding proteins.

\begin{figure*}[h!]
    \centering
\includegraphics[width=0.8\linewidth]{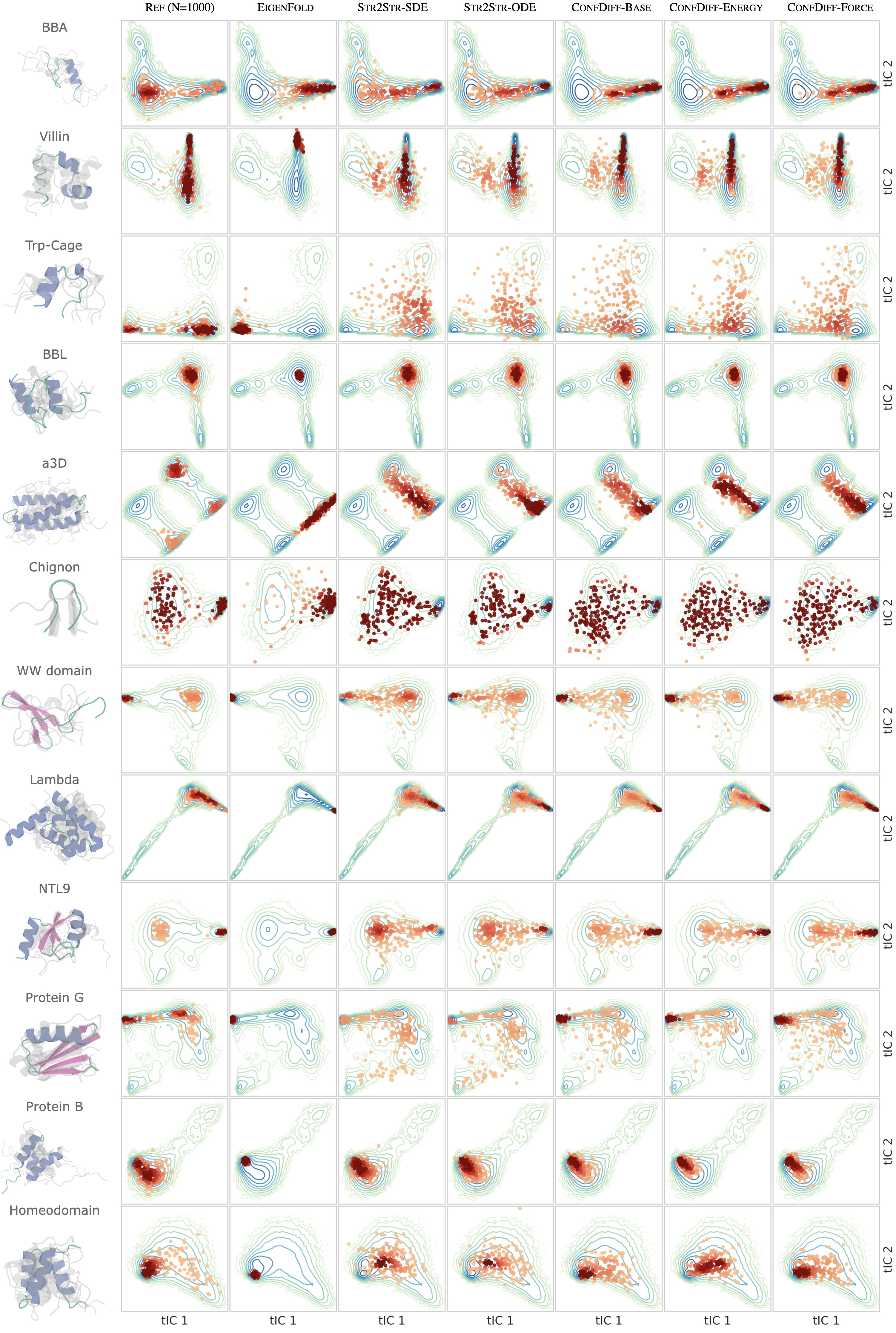}
    \caption{Sample distribution over the first two TIC components for twelve fast-folding proteins. Contours indicate the full reference MD sample density at log-scale.}
    \label{fig:fastfold-tica-proj-all}
\end{figure*}

\clearpage

\subsection{Additional Results for BPTI}

We list the best RMSD scores to each five cluster centers in Table \ref{tab:bpti-cluster-all}. \textsc{\model-Force} and \textsc{EigenFold} can most accurately predict the structures of metastable states. Especially, \textsc{\model-Force} out-performs all other models in predicting the remote cluster 3.

\begin{table*}[ht!]
\label{tab:bpti-cluster-all}
\caption{Detailed best RMSD to each cluster center. Unit: \si{\angstrom}.}
\vskip 0.15in
\setlength\extrarowheight{2pt}
\begin{center}
\begin{small}
\begin{sc}
\begin{tabular}{llllll}
\toprule
Model & Cluster 1 & Cluster 2 & Cluster 3 & Cluster 4 & Cluster 5 \\
\midrule
Ref (N=1000) & 0.79 & 0.79 & 2.19 & 1.02 & 0.74 \\
EigenFold & \textbf{0.86} & 1.69 & 2.45 & \textbf{1.34} & \textbf{0.99} \\
Str2Str-ODE & 1.85 & 1.94 & 2.46 & 2.13 & 1.9 \\
Str2Str-SDE & 2.04 & 2.01 & 2.80 & 2.34 & 2.12 \\
\model-Base & 1.02 & 1.72 & 2.38 & 1.5 & 1.12 \\
\model-Energy & 1.16 & 1.72 & 2.38 & 1.49 & 1.11 \\
\model-Force & \textbf{0.86} & \textbf{1.58} & \textbf{2.12} & 1.44 & 1.03 \\
\bottomrule
\end{tabular}
\end{sc}
\end{small}
\end{center}
\end{table*}

\begin{figure}[h]
    \centering
    \vskip 1in
    \includegraphics[width=0.8
    \linewidth]{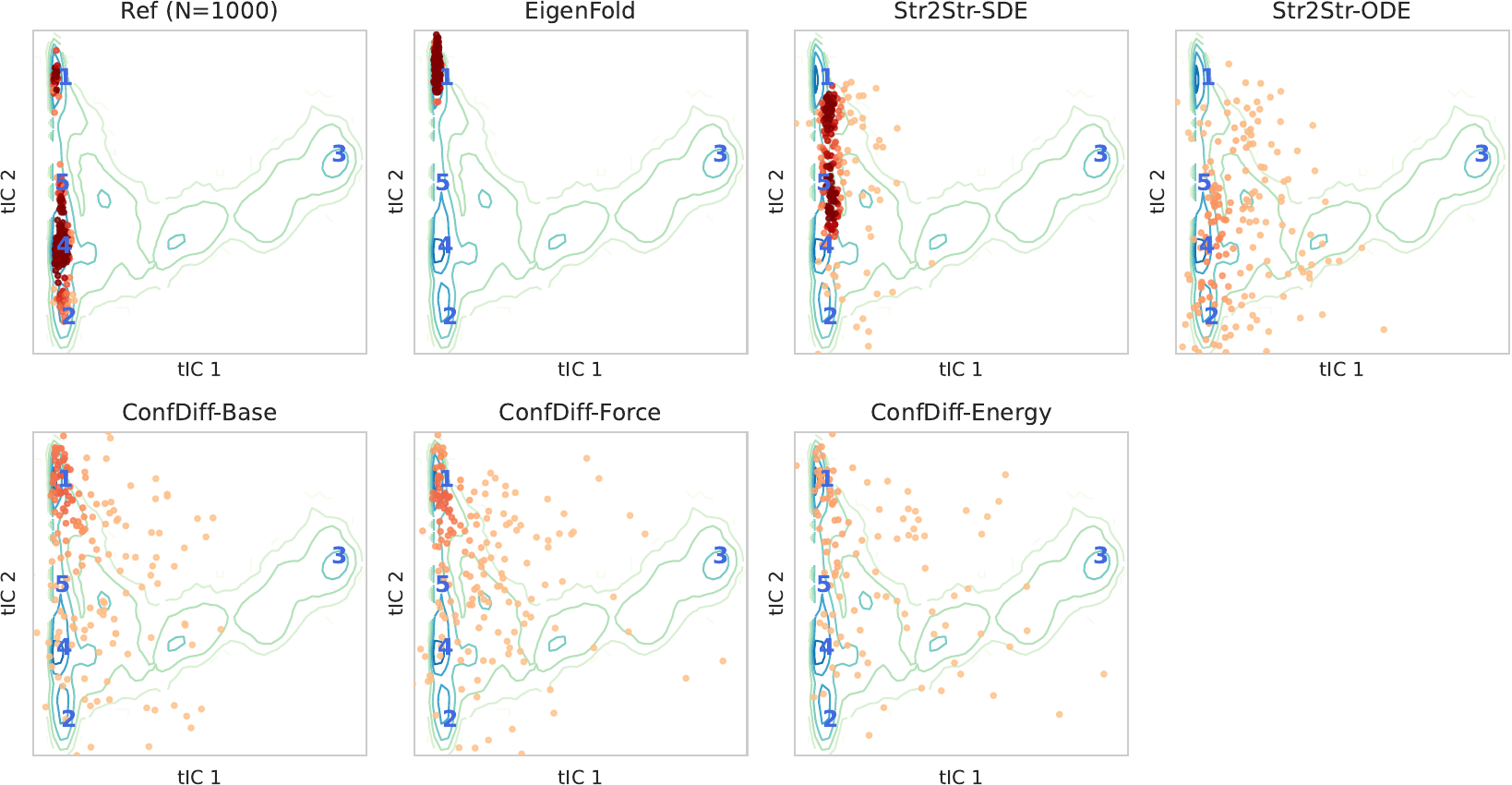}
    \caption{Model generated BPTI conformations projected to the first two TIC components. Contour lines represent the ground truth probability density at the log scale.}
    \label{fig:bpti-tica-proj}
\end{figure}

Similar to fast-folding, we visualize the sample distribution over the first two TIC components (Figure \ref{fig:bpti-tica-proj}). Despite that \textsc{EigenFold} predicts structures for cluster 2, 3, 4 better \textsc{Str2Str} and two \model models, its TICA plot shows it is concentrated around cluster 1 while other models have better sample coverage.

\clearpage
\subsection{\textsc{\model-Base} Results on Apo-holo Dataset}

\textbf{Apo-holo} is a dataset consists of 90 proteins with known conformation changes upon ligand-binding and each has two conformations (\textit{apo}, the unbound state; \textit{holo}, the bound state) determined by experiments. This dataset introduces a challenging task of predicting distinct conformations of proteins with various sizes and structural changes \cite{Eigenfold, saldano2022impact}. Due to the high computation cost of evaluating the energy and force for mid-to-large size structure using \texttt{OpenMM}, we only compare the performance of our \textit{classifier-free} model with baselines.

\begin{figure}[ht!]
    \vskip 0.5in
    \centering
    \includegraphics[width=0.8\linewidth]{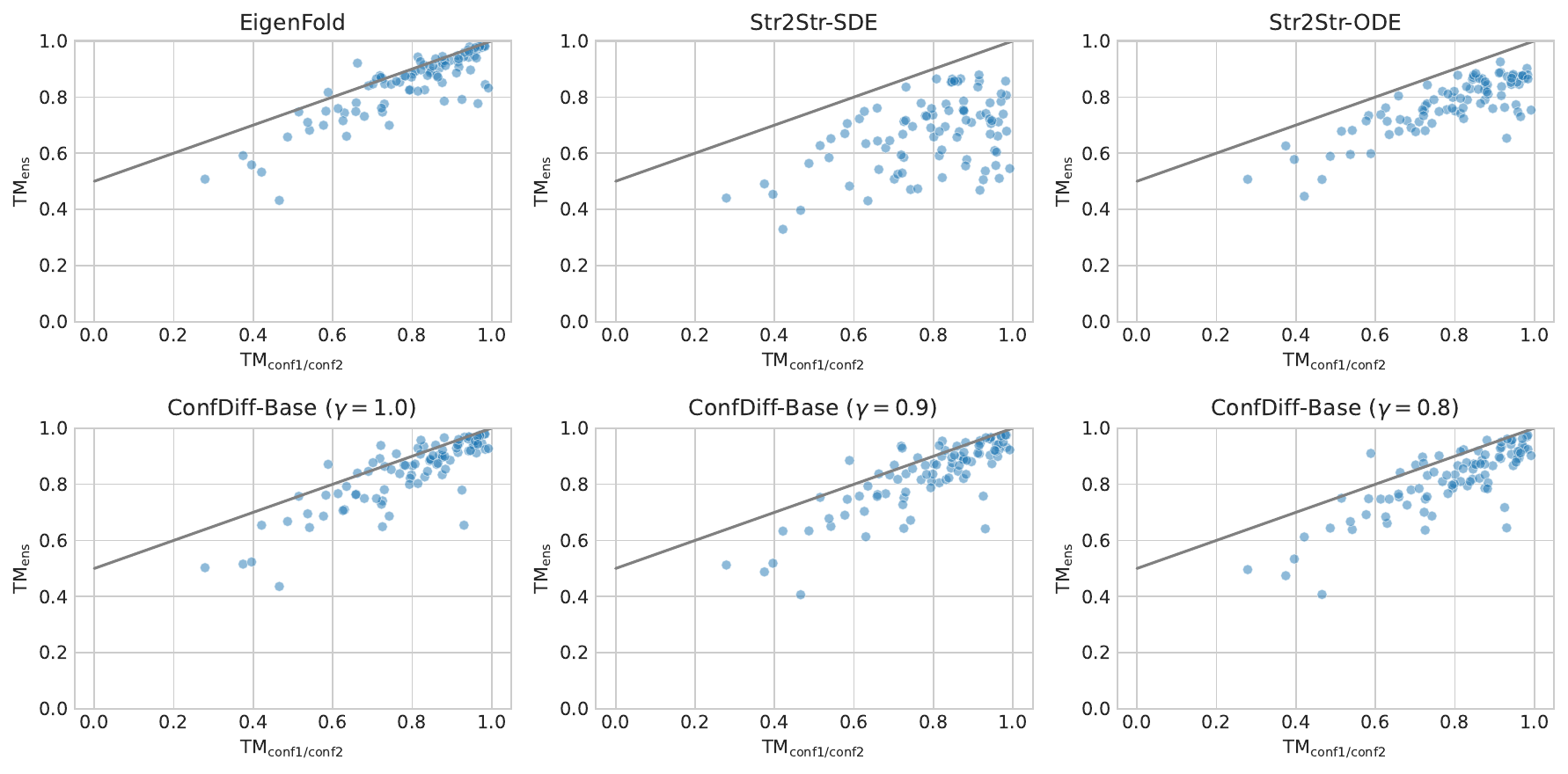}
    \caption{TMens vs TMscore between \textit{apo} and \textit{holo} structures. A small $t=0.1$ selected for \textsc{Str2Str} models. The slide line indicators the baseline performance of a perfect folding model.}
    \label{fig:apo-tmens}
\end{figure}

To evaluate the performance on predicting two distinct conformations, we calculated TM-ensemble score (\textsc{TMens}) as in \cite{Eigenfold, str2str}:

\begin{eqnarray*}
    \textsc{TMens} = 1/2 \times \Big (\max_i \big\{\text{TMscore}(\text{sample }i, \textit{apo}) \big\} + \max_i \big\{\text{TMscore}(\text{sample }i, \textit{holo}) \big\} \Big)
\end{eqnarray*}

where TMscore is a normalized structural similarity score between 0 and 1 \cite{tmscore}. Results are summarized in Table \ref{tab:apo} and Figure \ref{fig:apo-tmens}.  None of the model shows superb performance as most of the samples perform worse than a perfect folding model, indicating remaining challenges for current conformation models to sample distinct structures with potentially large conformation changes. Among models compared, \textsc{EigenFold} and \textsc{\model-Base ($\gamma = 1.0$)} show better performance that likely due to better prediction of a "folded" state for a given protein; while \textsc{Str2Str} models show lower performance potentially due to the lack of sequence information during perturbation for accurate folding prediction. 

\begin{table}[h!]
\caption{Apo-holo results (90 cases). \textsc{TM-Div} is the mean pairwise TMscore between samples.}
\label{tab:apo}
\vskip 0.15in
\begin{center}
\begin{small}
\begin{sc}
\begin{tabular}{lllll}
\toprule
 & TMens & RMSF & TM-div \\
\midrule
EigenFold & 0.846/0.875 & 3.44/1.19 & 0.903/0.958 \\
Str2Str-ODE & 0.775/0.790 & 1.92/1.94 & 0.894/0.907  \\
Str2Str-SDE & 0.668/0.684 & 4.81/4.70 & 0.677/0.663  \\
\model-Base ($\gamma = 1.0$) & 0.843/0.874 & 3.35/2.27 & 0.847/0.883 \\
\model-Base ($\gamma = 0.9$) & 0.834/0.869 & 4.12/2.91 & 0.790/0.813 \\
\model-Base ($\gamma = 0.8$) & 0.818/0.843 & 5.55/4.60 & 0.693/0.703 \\
\bottomrule
\end{tabular}
\end{sc}
\end{small}
\end{center}
\vskip -0.1in
\end{table}

\clearpage
\subsection{\textsc{\model} Results on Sars-CoV-2}
In order to fully understand the method's generalization capabilities, we did some proof-of-concept experiments to extend proposed method to challenging cases of large proteins - the SARS-CoV-2 case studies introduced in Distributional Graphormers (DiG) \citep{DiG}. This dataset contains two important proteins in drug discovery:  the spike receptor-binding proteins (RBD, 194 amino acids) and the main protease (Mpro, 306 amino acids). MD simulations for both proteins are publicly available and are used as reference during evaluation.

Preliminary results showed that protein sequence information using representations from ESMFold \citep{esmfold} is not able to predict reasonable structures of RBD and Mpro, while AlphaFold predictions are close to the experimental structures in Table \ref{fig:representation}.
\begin{table}[h]
\centering
\caption{The RMSD and TMscore of RBD and Mpro from different sequence representations.}
\vskip 0.1in
\begin{sc}
\begin{tabular}{ccc}
\toprule
 & RMSD & TMscore \\ 
\midrule
ESMFold\_RBD & 20.55 & 0.18 \\ 
ESMFold\_Mpro & 13.96 & 0.59 \\ 
AlphaFold\_RBD & 2.55 & 0.91 \\
AlphaFold\_Mpro & 3.28 & 0.93 \\ 
\bottomrule
\end{tabular}
\end{sc}
\label{fig:representation}
\end{table}

For quantitative comparison, we calculated the following metrics: 1) average RMSD to experimental structures (PDB IDs are RBD: 6LU7, Mpro: 6M0J); 2) the JS-TIC2D to reference distribution;  3) performances on the binary classification of TICA states similar to DiG: the joint space is discretized into $50 \times 50$ states and those with/without ground truth conformations are positive/nagtive states; 4) diversity as pairwise RMSD between samples (RMSF)  We summarized the metrics below in Table \ref{fig:sar-cov-2}, with some major findings:

\begin{table}[h]
\centering
\caption{Results on Sars-CoV-2. The results for classifier-free, energy guidance, and force guidance are labeled as \textsc{\model-Base}, \textsc{\model-Force}, and \textsc{\model-Energy}. All values are shown as mean.}
\vskip 0.1in
\label{fig:sar-cov-2}
\begin{sc}
\begin{tabular}{ccccccc}
\toprule
\multicolumn{1}{c}{} & \multicolumn{1}{c}{} & \multicolumn{1}{c}{} & \multicolumn{3}{c}{{Binary prediction of TIC2D states}} &  \\ \cmidrule(lr){4-6}
\multicolumn{1}{c}{\multirow{-2}{*}{{Model}}} & \multicolumn{1}{c}{\multirow{-2}{*}{{RMSD (Å)}}} & \multicolumn{1}{c}{\multirow{-2}{*}{{JS-TIC2D $\downarrow$}}} & \multicolumn{1}{c}{{Precision $\uparrow$}} & \multicolumn{1}{c}{{Recall $\uparrow$}} & \multicolumn{1}{c}{{F1 $\uparrow$}} & \multirow{-2}{*}{{RMSF (Å)}} \\ \midrule
\multicolumn{7}{c}{{RBD}} \\ \midrule
\multicolumn{1}{c}{\model-Base} & \multicolumn{1}{c}{{1.396}} & \multicolumn{1}{c}{\textbf{0.822}} & \multicolumn{1}{c}{{0.72}} & \multicolumn{1}{c}{{0.147}} & \multicolumn{1}{c}{{0.244}} & {1.55} \\ 
\multicolumn{1}{c}{\model-Energy} & \multicolumn{1}{c}{{1.341}} & \multicolumn{1}{c}{{0.826}} & \multicolumn{1}{c}{{0.647}} & \multicolumn{1}{c}{{0.133}} & \multicolumn{1}{c}{{0.221}} & {1.564} \\ 
\multicolumn{1}{c}{\model-Force} & \multicolumn{1}{c}{{1.385}} & \multicolumn{1}{c}{{0.825}} & \multicolumn{1}{c}{\textbf{0.765}} & \multicolumn{1}{c}{\textbf{0.154}} & \multicolumn{1}{c}{\textbf{0.257}} & {1.606} \\ \midrule
\multicolumn{7}{c}{{Mpro}} \\ \midrule
\multicolumn{1}{c}{\model-Base} & \multicolumn{1}{c}{{1.475}} & \multicolumn{1}{c}{{0.785}} & \multicolumn{1}{c}{1} & \multicolumn{1}{c}{{0.125}} & \multicolumn{1}{c}{{0.221}} & {1.692} \\ 
\multicolumn{1}{c}{\model-Energy} & \multicolumn{1}{c}{{1.448}} & \multicolumn{1}{c}{{0.794}} & \multicolumn{1}{c}{1} & \multicolumn{1}{c}{{0.116}} & \multicolumn{1}{c}{{0.208}} & {1.722} \\ 
\multicolumn{1}{c}{\model-Force} & \multicolumn{1}{c}{{1.559}} & \multicolumn{1}{c}{\textbf{0.768}} & \multicolumn{1}{c}{1} & \multicolumn{1}{c}{\textbf{0.132}} & \multicolumn{1}{c}{\textbf{0.234}} & {1.828} \\ 
\bottomrule
\end{tabular}
\end{sc}
\end{table}

\begin{itemize}[leftmargin=*]
    \setlength\itemsep{-0.1em}
    \item All the models can accurately predict the structure for both RBD and Mpro with mean RMSD$<$1.6 Å. The samples are scattered near the major peaks in the reference distribution as a single cluster. For RBD, the samples spread out in the same direction as the reference distributions.
    \item Adding energy or force guidance does not significantly impact the metrics. While \model-Force showed some positive signs on slightly improved metrics (e.g., JS-TIC2D for Mpro and Recall \& F1 scores for both proteins).
    \item The current prototype is only fine-tuned on the AF2 representation with two proteins (RBD and Mpro). Lack of training on large datasets could significantly impact model's performance. We could train the base model on the entire PDB structures with AF2 representations in the future and we believe the model's performance would greatly improve.
    \item Nevertheless, we aim to use these experiments to show that 1) proposed methods can scale up and apply to large proteins with real-world applications; 2) even with much simplification on experiments, the results showed some promising signs on proposed results.
\end{itemize}




\end{document}